\documentclass{aa}  
\usepackage{latexsym, graphicx, natbib}
\usepackage{txfonts}
\begin{document}
   \title{The Standing Wave Phenomenon in Radio Telescopes}

   \subtitle{Frequency Modulation of the WSRT Primary Beam}

   \author{A. Popping
          \inst{1, 2}
          \and
          R. Braun\inst{2}
          }

   \offprints{A. Popping}

   \institute{Kapteyn Astronomical Institute, P.O. Box 800, 9700 AV Groningen, the Netherlands\\
              \email{popping@astro.rug.nl}
         \and
             Australia Telescope National Facility, CSIRO, P.O. Box 76, Epping, NSW 1710, Australia\\
             }

   \date{}

 
  \abstract 
  {Inadequacies in the knowledge of the primary beam response of
 current interferometric arrays often form a limitation to the image
 fidelity, particularly when ``mosaicing'' over multiple telescope pointings.} 
 {We hope to overcome these limitations by constructing a
  frequency-resolved, full-polarization empirical model for the
  primary beam of the Westerbork Synthesis Radio Telescope (WSRT). }
  {Holographic observations, sampling angular scales between about~5
  arcmin and 11~degrees, were obtained of a bright compact source
  (3C147). These permitted measurement of voltage response patterns
  for seven of the fourteen telescopes in the array and allowed
  calculation of the mean cross-correlated power beam.  Good sampling
  of the main-lobe, near-in, and far-side-lobes out to a radius of
  more than 5 degrees was obtained.}
  {A robust empirical beam model was detemined in all polarization
  products (\textit{XX, XY, YX} and \textit{YY}) and at frequencies
  between 1322 and 1457~MHz with 1~MHz resolution. Substantial
  departures from axi-symmetry are apparent in the main-lobe as well
  as systematic differences between the polarization
  properties. Surprisingly, many beam properties are modulated at the
  5 to 10\% level with changing frequency. These include: (1) the main
  beam area, (2) the side-lobe to main-lobe power ratio, and (3) the
  effective telescope aperture. These semi-sinusoidsal modulations have a
  basic period of about 17~MHz, consistent with the natural ``standing
  wave'' period of a 8.75~m focal distance. The deduced frequency
  modulations of the beam pattern were verified in an independent long
  duration observation using compact continuum sources at very large
  off-axis distances.}  
  {Application of our frequency-resolved beam model should enable higher
  dynamic range and improved image fidelity for interferometric
  observations in complex fields, although at the expense of an increased
  computational load. The beam modulation with frequency can not be as
  easily overcome in total power observations. In that case it may prove
  effective to combat the underlying multi-path interference by coating all
  shadowed telescope surfaces with a broad-band isotropic scattering treatment.}


   \maketitle
%

\section{Introduction}
For any telescope it is essential to know the shape of the point
spread function as well as the effective field-of-view, since we can
only see the true sky as filtered by these response patterns. In the
case of an earth-rotation-synthesis interferometric array, the
effective field-of-view is determined by the complex product of each
pair of voltage response patterns of the telescopes in that
array. However, for most synthesis arrays, the precise shape of this
so-called ``primary beam'' is not known. For example the primary beam
of the 25~m prime-focus-fed parabolic dishes of the Westerbork
Synthesis Radio Telescope (WSRT) is approximated by the $\cos^6(c \nu
r)$ analytic function. In this function, $r$ is the distance from the
pointing center in degrees, $\nu$ the observing frequency in GHz and
the constant $c=68$ is, to first order, wavelength independent at GHz
frequencies.

This approximation assumes an axi-symmetric beam shape which consists
of just the main lobe and does not include any side-lobes.  For real
radio telescopes however, like the WSRT, the sensitivity is not
confined to the main beam, but spread out over the full $4\pi$
rad$^2$.  The telescopes are primarily sensitive in the main beam, but
a significant contribution comes from outside the main beam. Response
from outside the main beam is referred to as \textit{side lobes} or
\textit{stray radiation } \citep{1996A&AS..119..115H}. These
features can be caused by parts of the aperture which are blocked by
the feed and support structures \citep{1999ASPC..180...37N} as well as
by irregularities of the reflector surface. Radiation may be received
directly into the feed, but also after scattering off the feed support
legs.  \cite{1962dnwo.book.....V} was the first observer to interpret
temporal variations in \ion{H}{i} spectra as due to side lobes or stray
radiation. \cite{1980A&A....82..275K} documented stray radiation
problems for the Effelsberg telescope. Similar investigations have
been made for the Dwingeloo 25 meter telescope by
\cite{1962dnwo.book.....V}, by \cite{1962BAN....16..321V}, and by
\cite{1966BANS....1...33R}. From these investigations it is obvious
that essentially all observations made with parabolic radio antennae
are affected by stray radiation at some level.

Another phenomenon plaguing spectroscopic observations in radio
astronomy is the so-called ``standing wave'' phenomenon
\citep{1997PASA...14...37B}. The term ``standing waves'' is used to
describe a semi-periodic variation in the spectral bandpass seen in
many radio telescopes which has a basic wavelength of approximately
$c/(2f)$ Hz, for focal distance, $f$, (or about 17 MHz for the WSRT
dishes). As we will show below, this phenomenon is closely tied to
variations in the primary beam properties with frequency.

The antenna pattern of the 25~m telescope in Dwingeloo has been
studied in detail by \cite{1967BANS....2...59H}. So-called ``stray
cones'' are associated with radiation scattered from the feed support
legs, while a ``spillover ring'' is associated with radiation reaching
the feed from just beyond the edge of the
reflector. \cite{1980A&A....82..275K} measured the antenna pattern of
the 100~m Effelsberg telescope to a radius of 2 degrees from the main
beam axis and created an empirical model of the far side
lobes. Observations were done several times at different hour angles
and at various epochs, which provided the material necessary to
determine the relative amplitudes of the various components of the
model. The empirical model Kalberla developed contained a spillover ring,
four stray cones (from radiation scattered off the feed support legs),
and a blockage (or ``shadowing'') contribution due to the support
legs. Reflections from the roof of the apex cabin resulted in four
small additional components. With this work, Kalberla demonstrated
that the main beam and side lobes could be effectively modelled, using
careful measurements of the sky brightness distribution.

The WSRT is comparable to many other radio telescopes in the sense
that it has feed- and feed-support structure blockages, which
influence the shape of the beam. Departures from axi-symmetry and
side-lobes are present, but are neglected in the current beam
approximation. The approximation is truncated above the level of the
first side lobes, which directly limits the fidelity of a
primary-beam-corrected image. Image fidelity could be substantially
improved with a better model. A related problem arises when one is
combining data obtained at a series of different pointing centers, in
a process termed mosaicing \citep{1989ASPC....6..277C}, since an
accurate model of the primary beam is required when undertaking a
joint deconvolution of all data. The dynamic range in the resulting
combined image is determined (amongst other things) by the quality of
the primary beam model.

Although a simple approximation is used for most radio
telescopes, a specific detailed model would need to be
developed for each individually. We are particularly interested in
utilizing an improved beam model for the
WSRT and will develop that model here. The WSRT is an
aperture synthesis interferometer that consists of a linear array of
14 antennas arranged on a 2.7 km East-West line. Ten of the telescopes
are on fixed mountings while the remaining four dishes are movable
along two rail-tracks. The antennas are equatorially mounted 25-m
dishes (with an $f/D$ ratio of 0.35) and are fed from the prime
focus. A major advantage of the equatorial mounting is that the
primary beam does not rotate on the sky while tracking
\citep{1999ASPC..180...37N} as it does for the elevation-over-azimuth
mount \citep{1986isra.book.....T}.

The WSRT array is equipped with Multi Frequency Front Ends (MFFE),
which have cooled receivers at 3.6, 6, 13, 18+21 cm, and uncooled
receivers at 49 cm and 92 cm. The 21 cm receiver is used most
extensively and in this paper we will only consider beam properties in
this wavelength band. We develop a detailed numerical representation
of the WSRT-primary beam; not just including the main lobe, but also
the side lobes. Furthermore, we consider all four polarization
products individually rather than just Stokes I. And most importantly
we consider the variation of beam properties with frequency at a
resolution sufficiently fine (1 MHz) to fully sample the so-called
``standing wave'' phenomenon. The paper is organized as follows. In
section \ref{theory} we briefly describe the relationship between
aperture illumination and the beam pattern of a telescope. Section
\ref{method} describes the method used to determine our model and in
section \ref{results} the results are presented. The model is tested
with observations of celestial sources in section \ref{testing}. In
the summary and discussion of section \ref{summary} we consider how
frequency modulation of beam parameters can be dealt with, both for
interferometric and total power observations.


\section{Theory}
\label{theory}
We will briefly consider the relationship between aperture
illumination and a radio telescope beam. This will be useful when we
describe how we arrive at our beam representation in the next section.
The relation between the complex aperture pattern of the antenna,
$f(u,v)$, and the complex voltage-beam pattern, $F(l,m)$, of the
antenna \citep{1986raas.book.....K} is:
\begin{equation}
F(l,m) = \int\!\!\!\int_{aperture} f(u,v)e^{2\pi i(ul+vm)} \mathrm{d}u \mathrm{d}v
\end{equation}
and
\begin{equation}
f(u,v) = \int^{\infty}_{-\infty}\int^{\infty}_{-\infty} F(l,m)e^{-2\pi i(ul+vm)} \mathrm{d}l \mathrm{d}m
\end{equation}
The coordinates of the radiation pattern are given by
\begin{equation}
u = \sin \theta \cos \phi \textrm{ and } v = \sin \theta \sin \phi
\end{equation}
This Fourier transform relationship between antenna and voltage-beam
pattern is analogous to the one between the source brightness
distribution and the visibility function that is sampled by an
interferometer. The aperture distribution $f(u,v)$, is determined by the way in
which the antenna feed illuminates the aperture together with how the
aperture is blocked, as is described in detail by
\cite{1999ASPC..180...37N}. In general, the more that $f(u,v)$ is
tapered at the edge of the aperture, the lower will be the aperture
efficiency and the side lobes and the broader the main beam. The
Fourier transform of the aperture can be described in terms of
\textit{amplitude} and \textit{phase} or the real and imaginary parts,
$(a + ib)$. In the case of a single dish telescope or when calculating
auto-correlations, the power patterns are given by $F \cdot F^*$,
which gives
\begin{equation}
(a + ib) \times (a - ib) = a^2 + b^2
\end{equation}
In the case of cross-correlations, when using different telescopes,
the power pattern is given by $(F \cdot G^* + G \cdot F^*) / 2$, which
gives
\begin{equation}
\frac{(a + ib) \times (c - id) + (c + ib) \times (a - ib)}{2} = ac + bd
\end{equation}

\begin{figure*}[!t]
\includegraphics[width=0.33\textwidth]{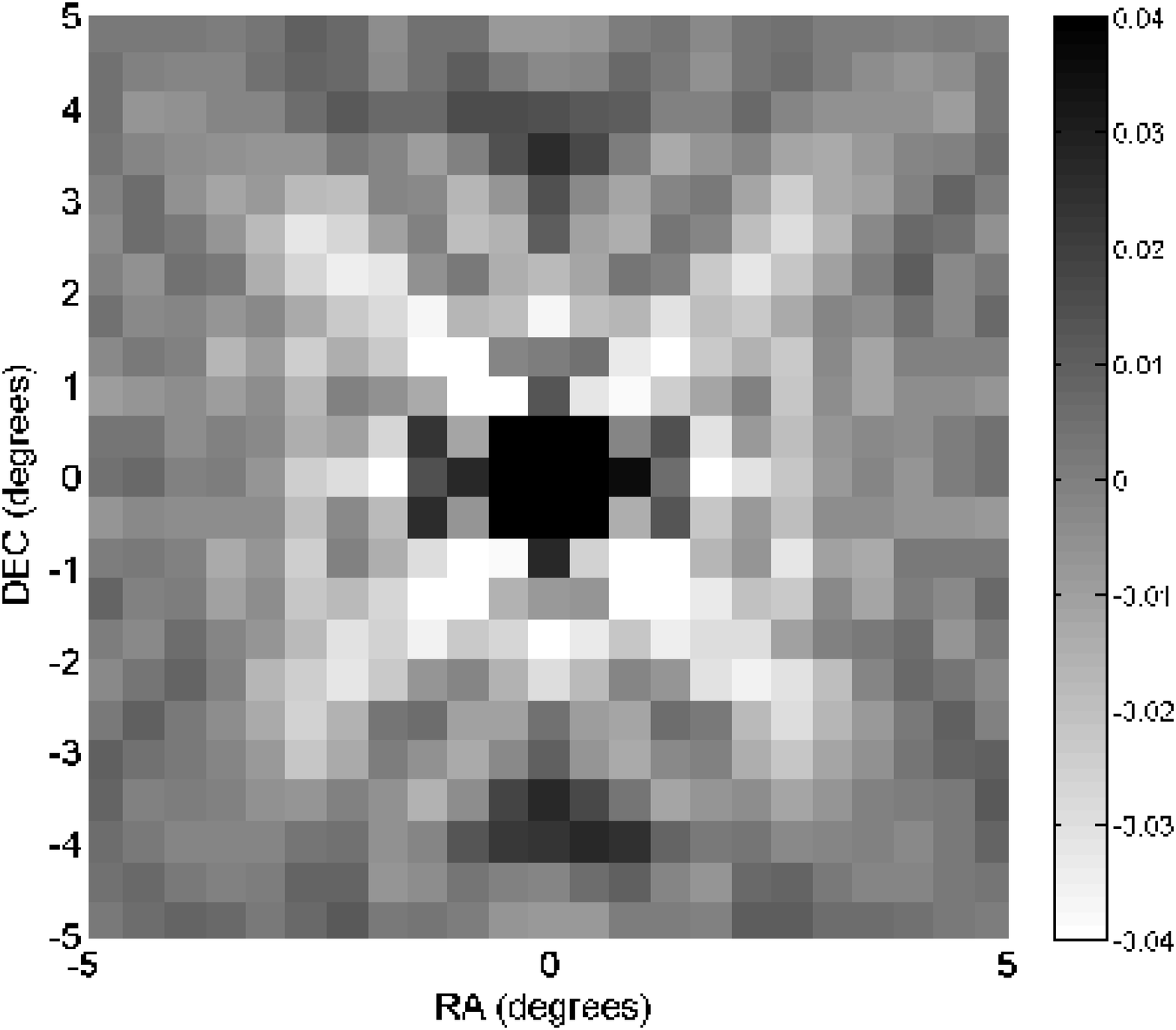}
\includegraphics[width=0.33\textwidth]{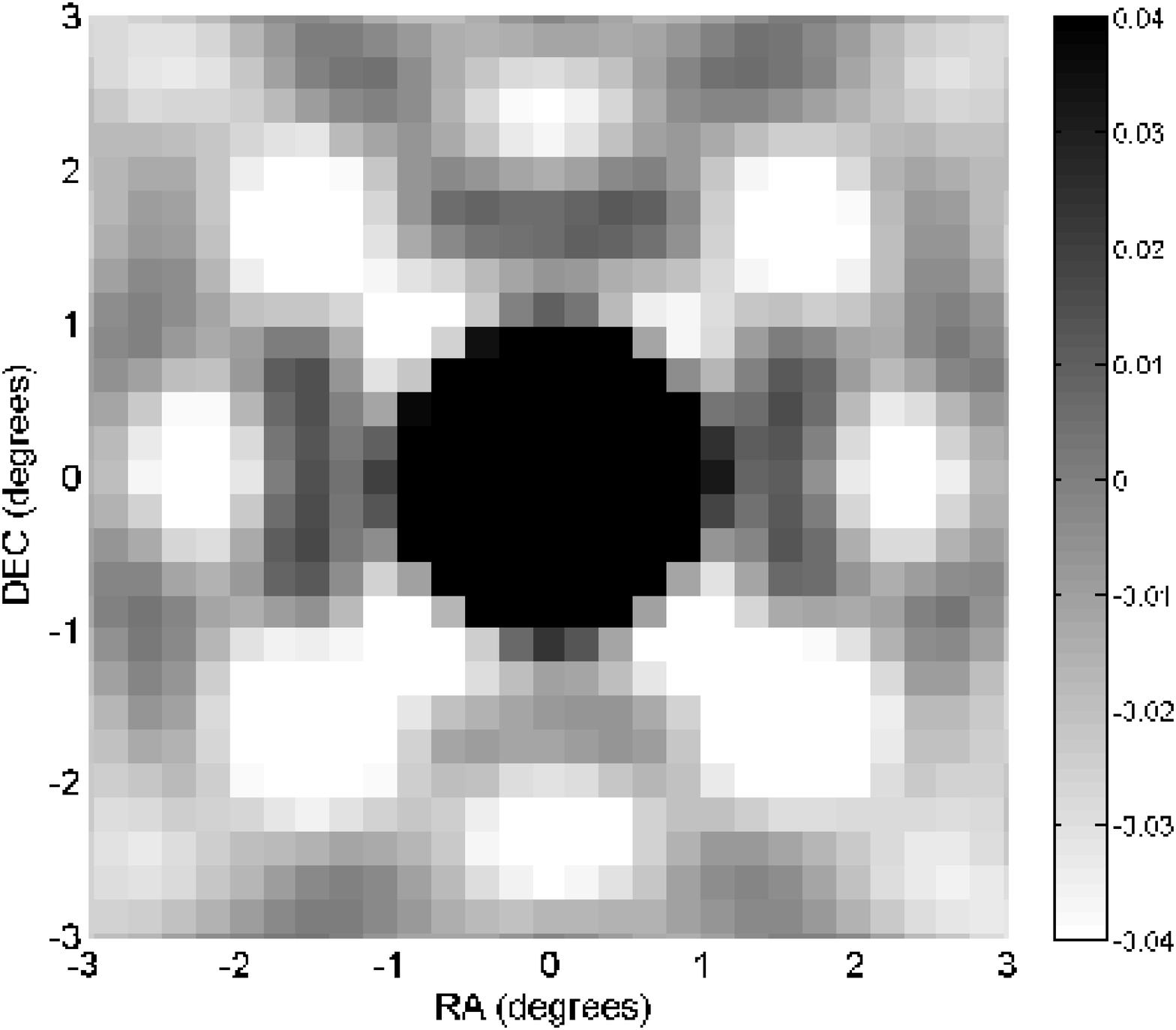}
\includegraphics[width=0.33\textwidth]{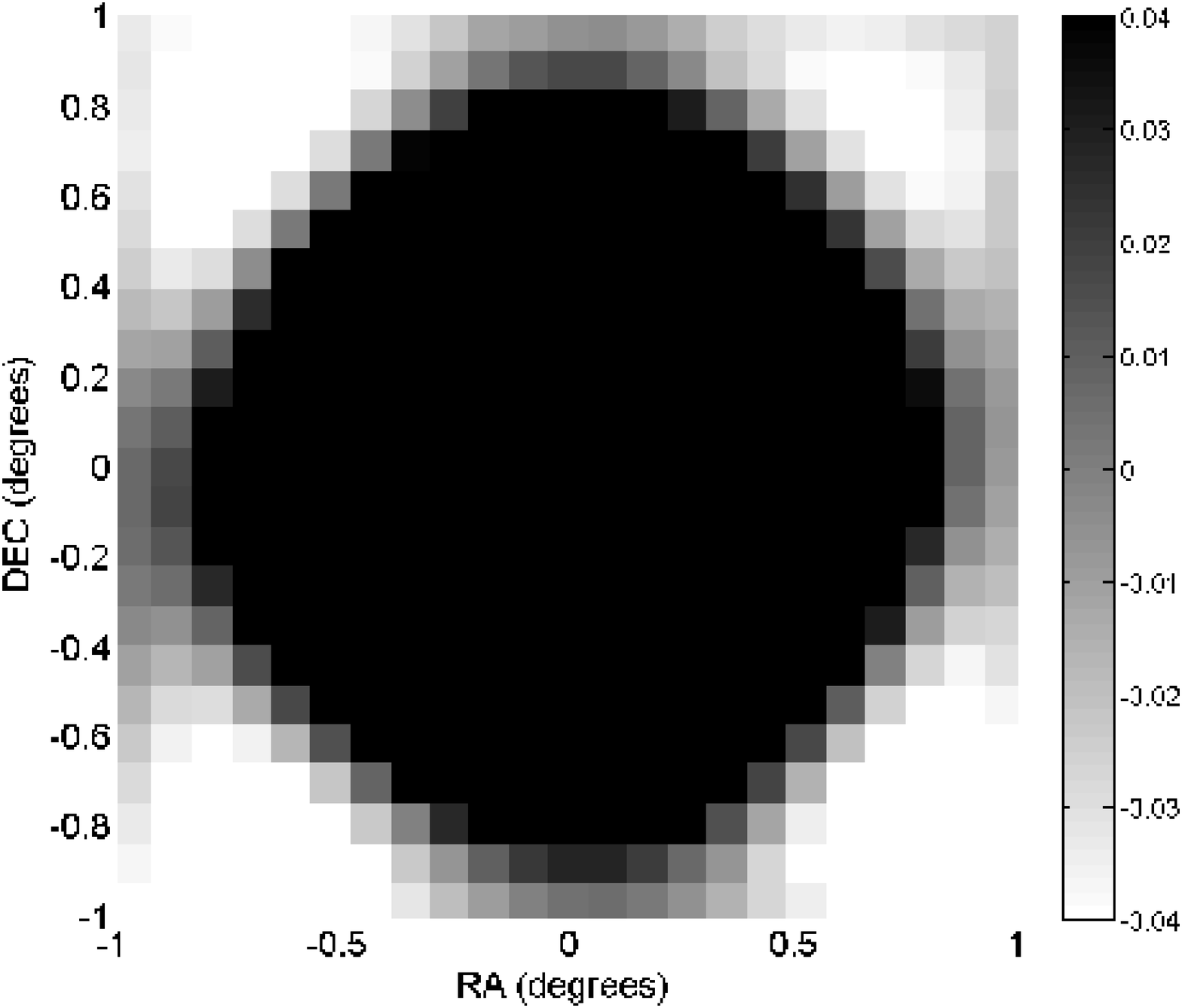}
\includegraphics[width=0.33\textwidth]{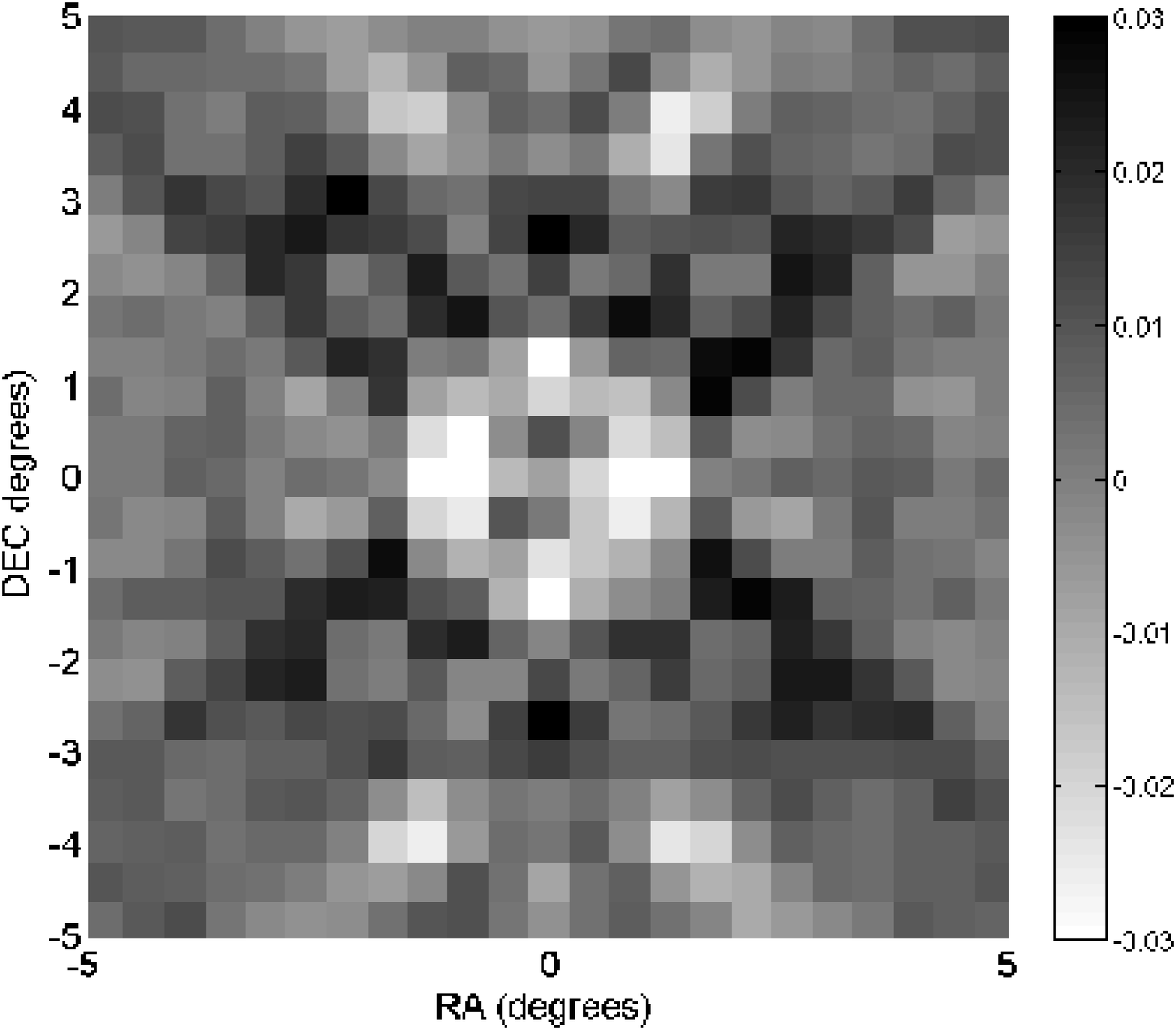}
\includegraphics[width=0.33\textwidth]{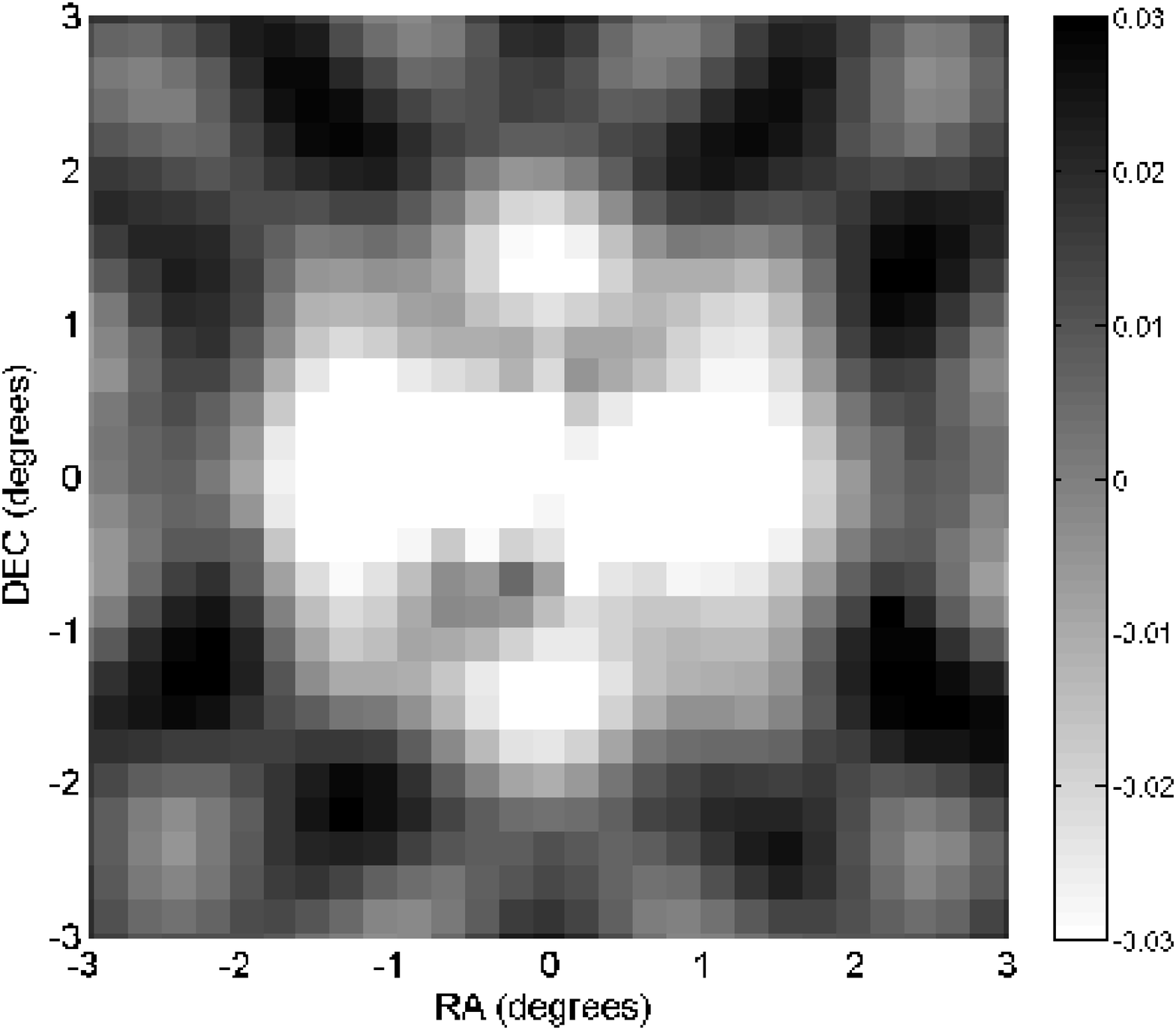}
\includegraphics[width=0.33\textwidth]{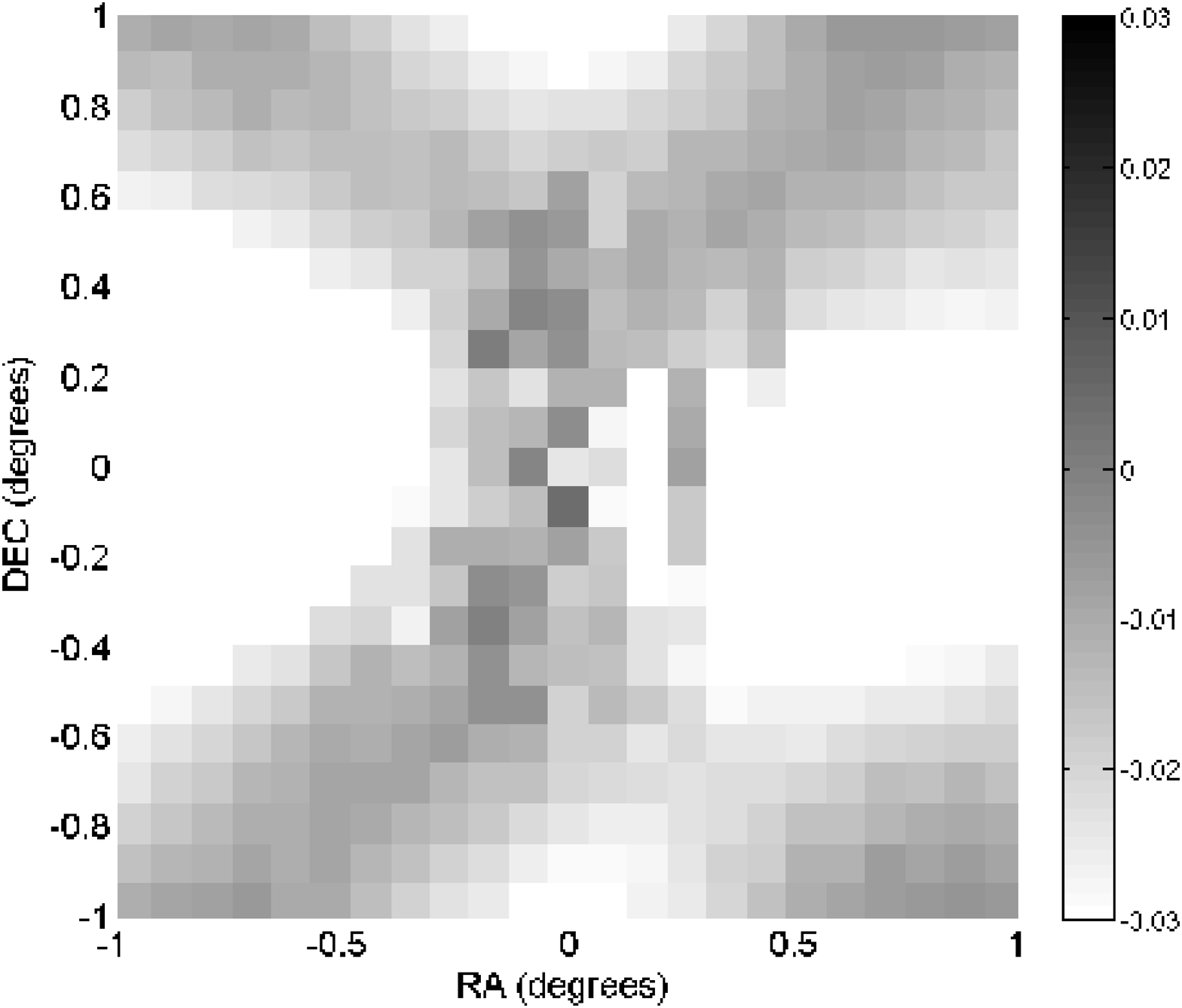}
\caption{PSFs of the voltage pattern of polarisation $X$ at 1416.6 MHz
for the three different holographic measurement scales, averaged over
all the antennas. From left to right is the large, medium and small
field of view. The top panels show the {\it cosine} (real) values
and the bottom panels show the corresponding {\it sine} (imaginary)
values. All images are on a linear scale, although for the real images
only a small range is plotted at the level of the sidelobes. The peak
values of the voltage patterns are normalised to unity.  }
\label{voltage}
\end{figure*} 

\begin{figure*}[!t]
\includegraphics[width=0.33\textwidth]{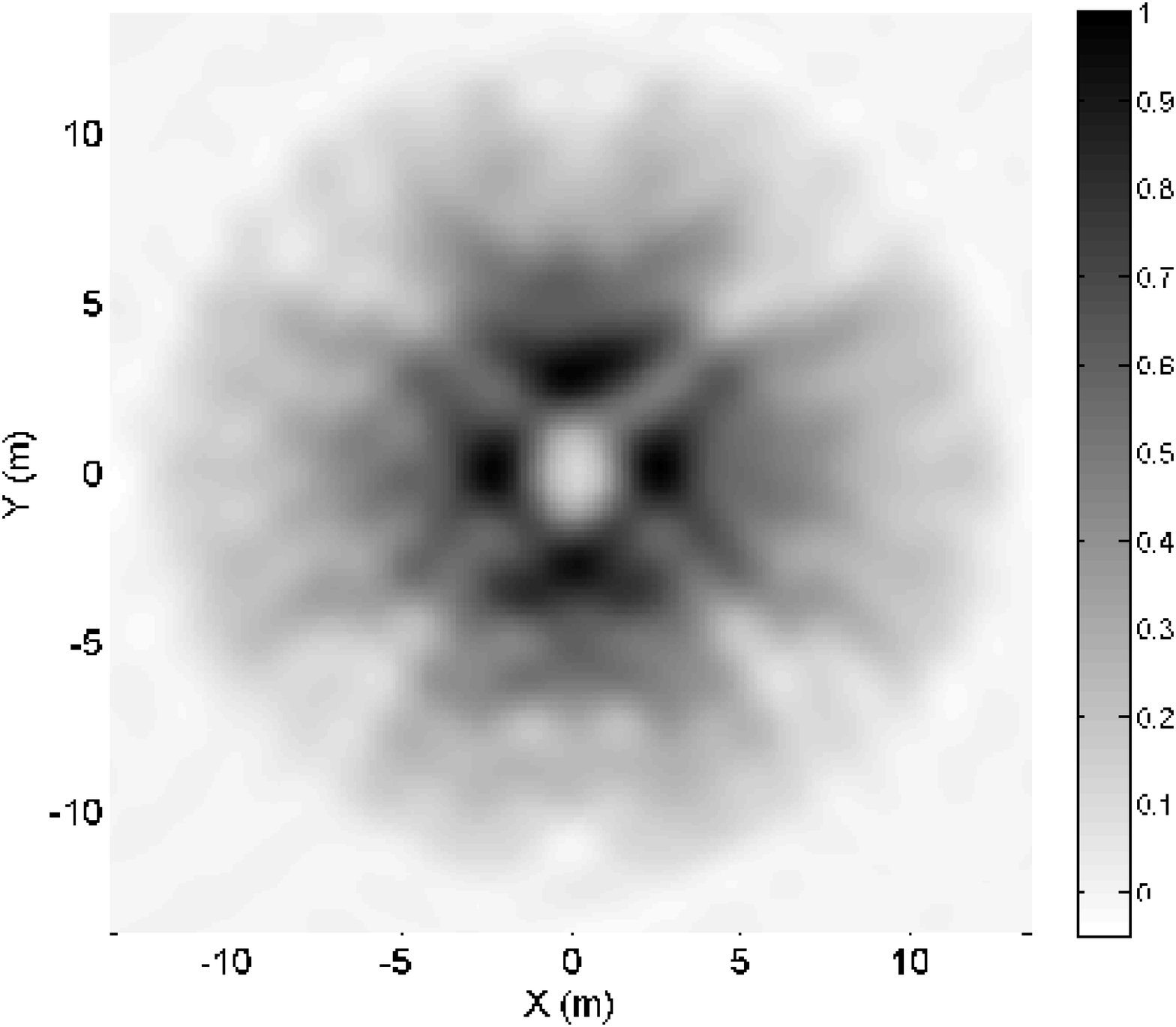}
\includegraphics[width=0.33\textwidth]{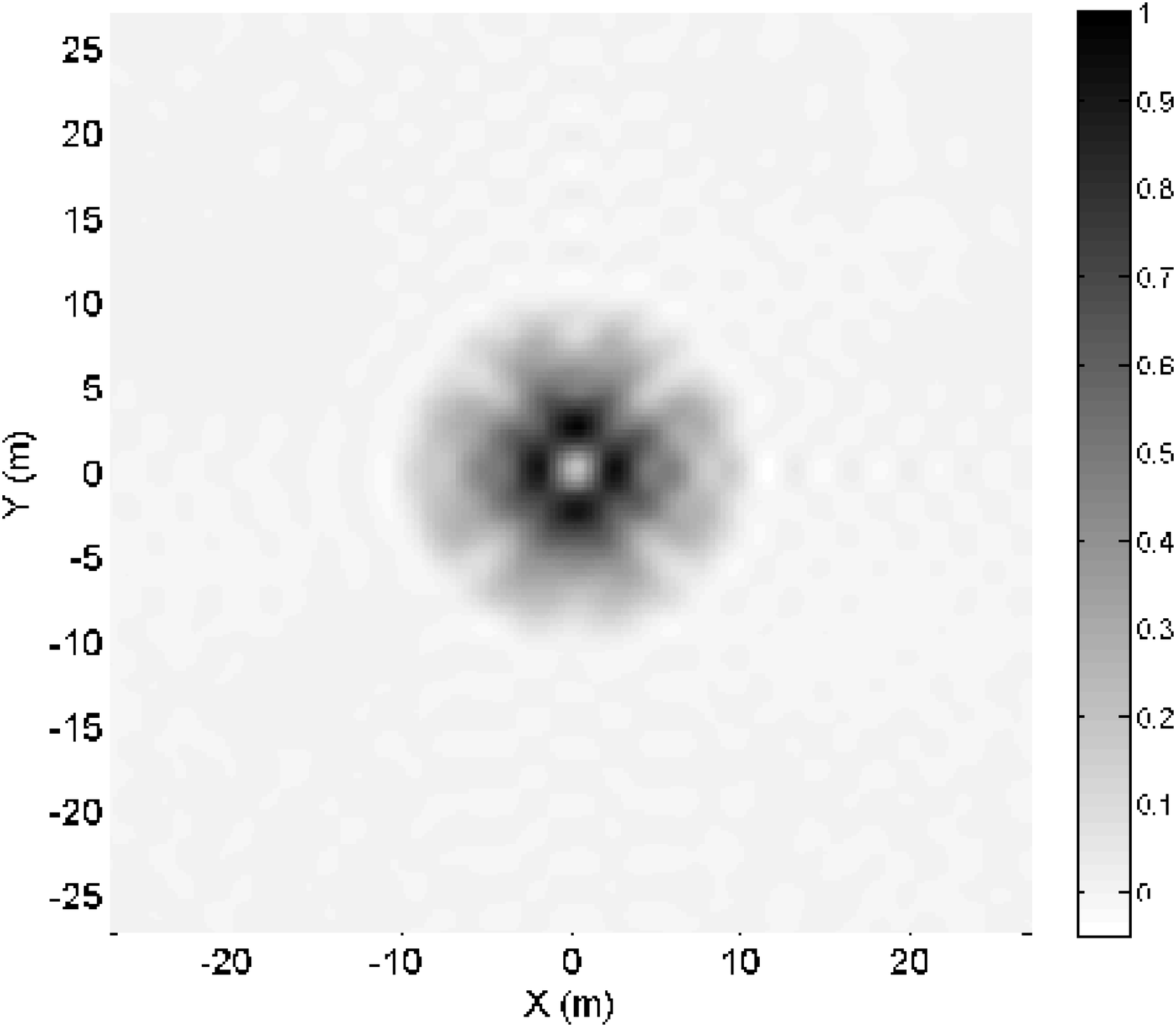}
\includegraphics[width=0.33\textwidth]{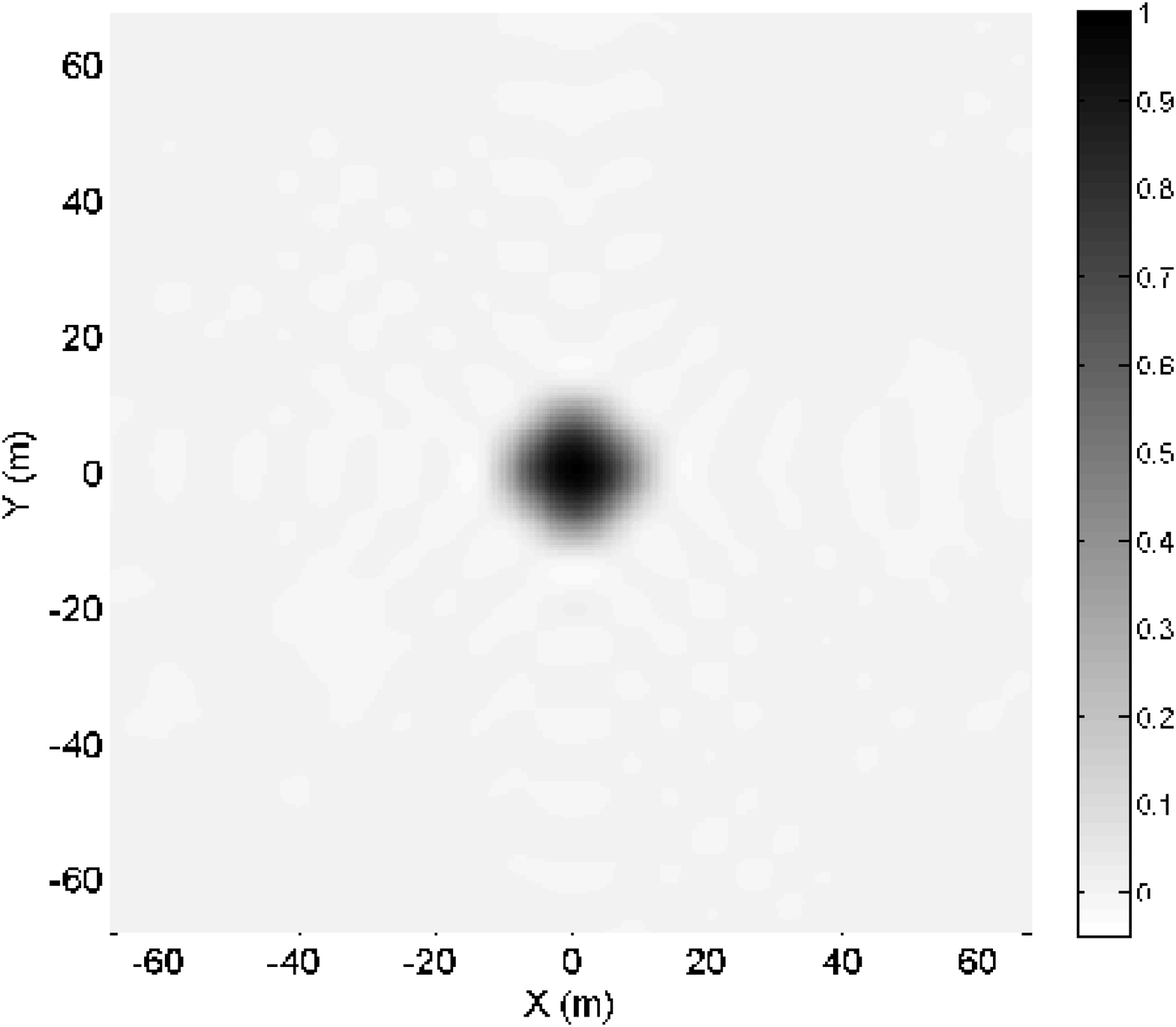}
\includegraphics[width=0.33\textwidth]{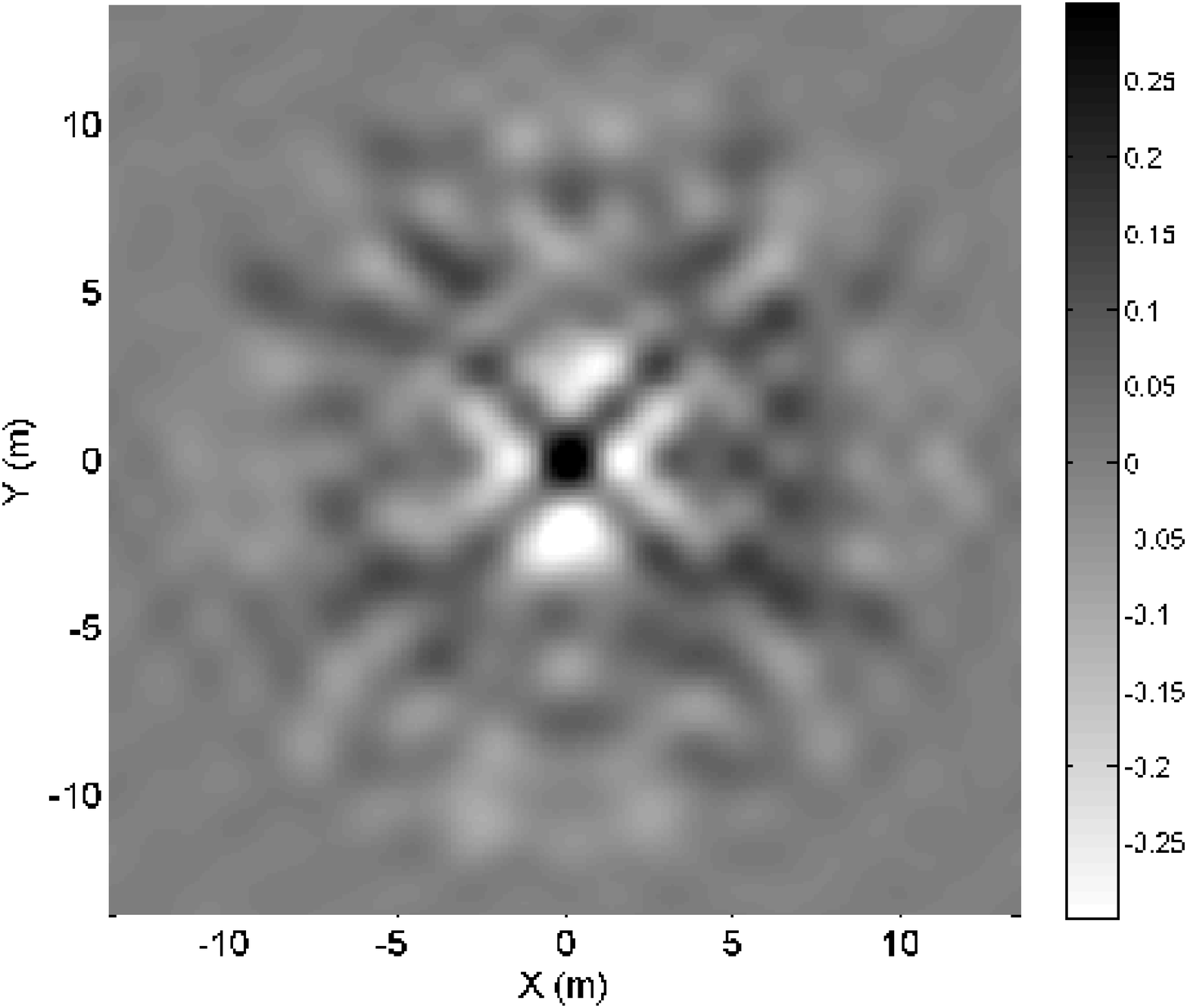}
\includegraphics[width=0.33\textwidth]{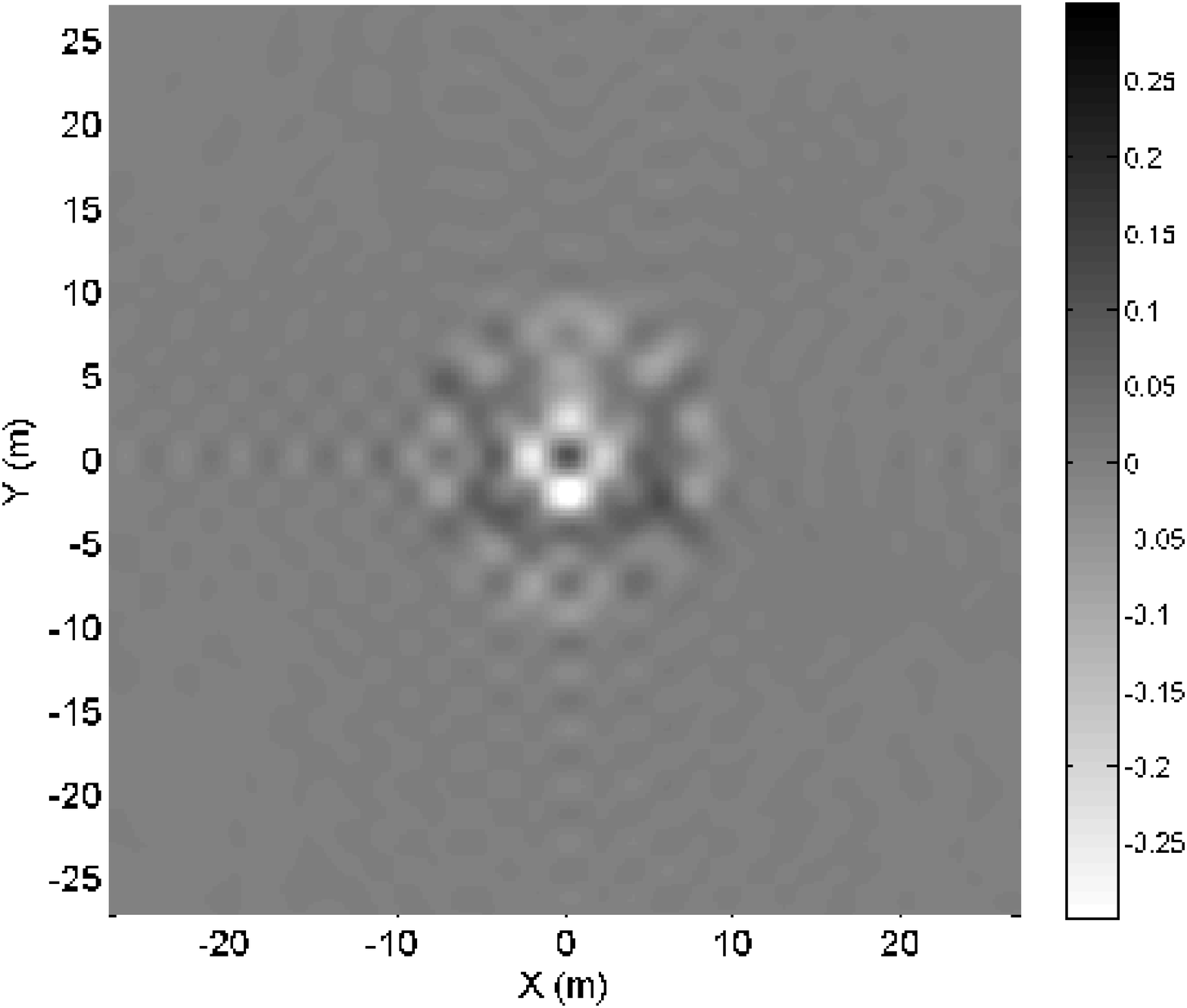}
\includegraphics[width=0.33\textwidth]{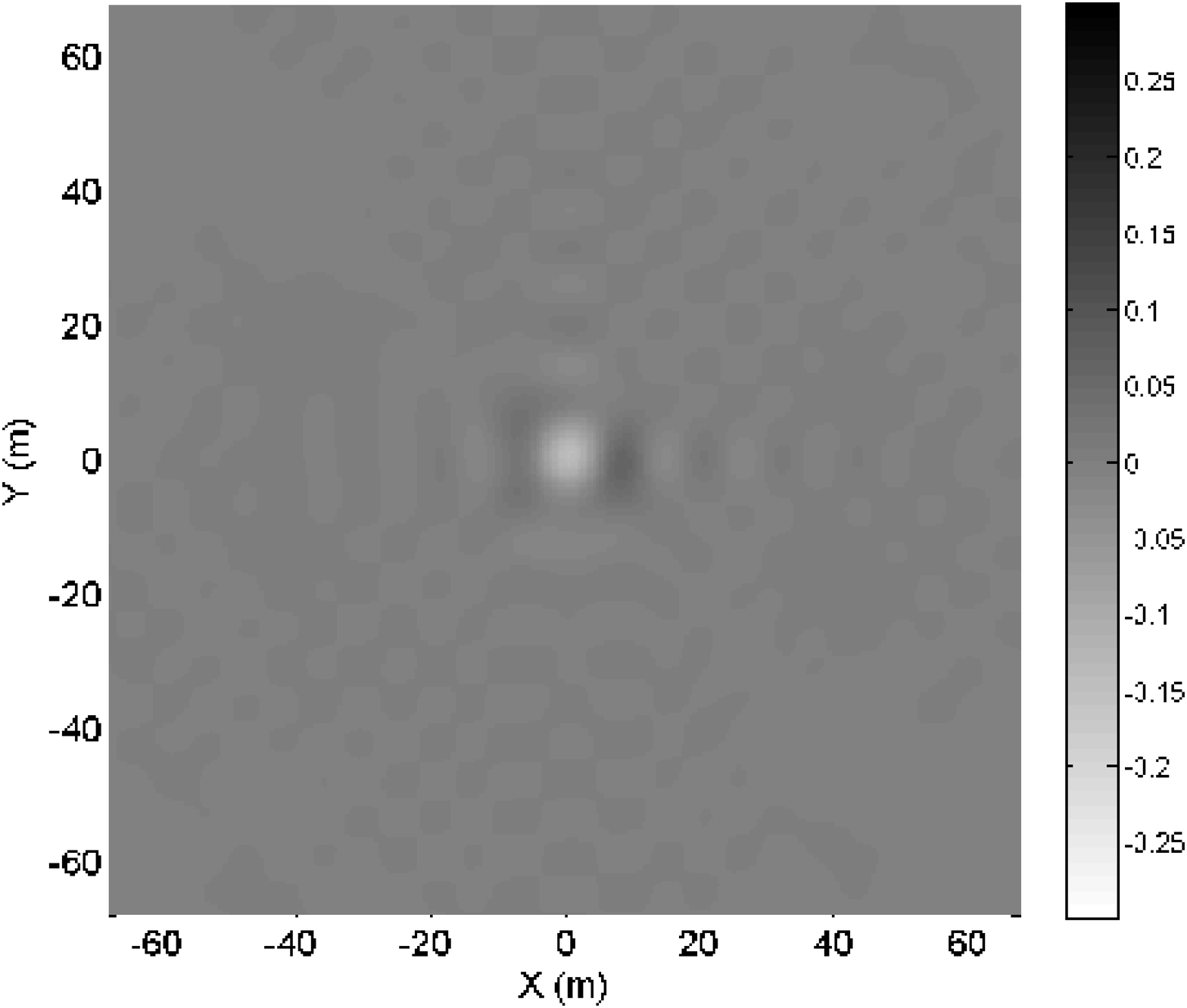}
\caption{Aperture patterns corresponding to the voltage patterns in
Fig.~\ref{voltage}. The large field of view gives the most detailed
description of the antenna, showing clearly the regions blocked by the
receiver and support structure. Phase deviations are largest within
the blocked regions of the aperture.  }
\label{aperture}
\end{figure*}

\section{Method}
\label{method}

An analytic model would be the most convenient way to describe the
beam pattern. This could be used to predict the beam parameters at any
desired position without the need for interpolation and would be easy
to implement in software. However as we will show below, an analytic
model is not sufficient to describe the actual telescope beams and an
empirical model is required. A straightforward way to get the actual
beam shape of a telescope is by measuring it's voltage pattern with a
holographic measurement (\cite{1977MNRAS.178..539S};
\cite{1986raas.book.....K}). A pair of antennas is needed; whereby one
of the antennas continuously tracks a strong point source and the
other antenna scans the vicinity of that source. The
voltage pattern $F(l,m)$ is obtained from the complex correlation
coefficient of the scanning antenna with respect to the reference
antenna. Once the voltage pattern is known the auto-correlation power
beam can be obtained by multiplying with the complex
conjugate. Alternatively, the interferometric cross-correlation power
beam can be obtained by multiplying voltage patterns and their complex
conjugates in a pair-wise fashion and averaging the result. Our model
is based on this concept of holographic measurements, although it is
extended by combining a series of measurements that sample a wide
range of angular scales.

\subsection{Holographic Measurements}
Three holographic observations were carried out, with different fields
of view to obtain sampling at high, medium and low resolution. The
smallest field of view sampled 2.2$\times$2.2 degrees with
25$\times$25 samples, the medium, 7.26$\times$7.26 degress with
33$\times$33 samples and the large, 11$\times$11 degrees with
25$\times$25 samples. For each observation 8 bands of 20 MHz bandwidth
were observed with 8$\times$64 spectral channels and the 4 linear
polarization products: {\it XX}, {\it XY}, {\it YX} and {\it YY}. Of
the 20 MHz of each band, about 18 MHz are useful due to the roll-off
of the intermediate frequency filters at the edge of the band. The
bands were centered at 1450, 1432, 1410, 1392, 1370, 1350, 1330, 1311
MHz. The frequencies were selected to avoid Galactic neutral hydrogen
emission and known radio frequency interference (RFI eg. near 1380
MHz).

Of the 14 telescopes in the array, the seven even-numbered telescopes
were used as reference antennas and the seven odd-numbered were used
for scanning the vicinity of 3C147. A 10 second basic integration time
was employed, with 30 seconds nominal dwell time on each position in
the scanning pattern. Data acquired during telescope movement was
flagged, yielding 20 seconds of net integration time per position. The
voltage beams were obtained from the holographic measurements using
the MSHOLOG code developed by H. van Someren Greve. This code
calibrates out constant phase offsets between antennas and compensates
for the small geometrical delays that arise when scanning telescopes
are pointed off-axis. An optional Wiener filtering can be employed
within MSHOLOG to limit the occurence of signal in the aperture plane
to only the surface of the reflector, while interpolating in the
far-field sampling pattern. We have not employed this option, but
chose only to grid the calibrated data. Each voltage beam has a real
and imaginary part, which can be recast as amplitude and phase. An
important choice in the processing is the number of spectral channels
to bin together. Binning more channels improves the signal-to-noise
ratio, but degrades the spectral resolution. We chose to bin each
group of two consecutive channels, yielding 0.625 MHz spectral
resolution. Individual voltage beam images were inspected and an
occasional grid-point was replaced with a linearly interpolated value
when large deviations were apparent. In the handful of cases where
excessive deviations were encountered
in the entire pattern, i.e. when the data quality was not sufficient,
that specific pattern was not included in the subsequent processing.

The pair-wise (complex conjugate) products of the seven available
voltage beams yielded 21 distinct baseline power beams. These were
averaged to yield the mean interferometric power beam. All power beams
were normalised to a peak height of unity. The average dispersion
of a single combination of antennas with respect to the mean power
beam is $~0.2\%$. This number is determined by calculating the
dispersion of each combination at each frequency and taking the mean
value of these individual dispersions. For this purpose we only made
use of the holographic measurements with a small field of view,
mapping the main beam at high resolution.

\subsection{Combining Measurements} 

The mean power beams at three different scales need to be combined to
get the most complete representation of the beam. The small field
provides detailed information about the center of the beam, the medium
field samples intermediate scales, while the large field samples the
outer side lobes.

To combine the images, the sampling of the three measurements should
ideally be the same. This can be accomplished with a {\it Sinc}
function interpolation by Fourier transforming the medium and large
field data to the aperture plane and padding out in the transform
plane with zeros before Fourier transforming back to the image
plane. The sampling of our small field was chosen to be exactly five
times as fine as our large field and 2.5 times as fine as the medium
field. So padding the Fourier transform of our medium field out to 2.5
times the size and our large field to five times the image size before
Fourier transforming back yields identical image plane sampling for
all three power beams. The difference is only the spatial extent
provided by each, which has not changed. Beginning with the resampled
large beam, the inner 7.26$\times$7.26 degrees were first replaced
with the resampled medium beam and finally the inner 2.2$\times$2.2
degrees were replaced with the measured small beam. Minor
discontinuities between the different coverages were smoothed out by
taking the average within a 0.25 degree wide border zone around the
medium and small beams rather than taking the simple replacement as
done inside this zone.

It is quite instructive to consider both the image and aperture plane
features that are apparent on the three observed scales.  An example
of the three observed voltage beam scales is shown in
Fig.~\ref{voltage}, while the corresponding aperture distributions are
shown in Fig.~\ref{aperture}. The large field image has an aperture
distribution which is only slightly larger than the size of the
telescope surface. The resolution is moderately high, yielding a
detailed view of the shadowed parts of the telescope surface. The
feed-support legs can be clearly seen as well as the triangular
shadows that these legs cast on the outer reflector surface. The top
right support leg has a slightly larger electrical cross-section than
the other three since the cables from the receiver package also run
along it. Remarkably, there is significant signal intensity from the
blocked parts of the aperture; for example under the
receiver. Geometric optics would preclude signal intensity from such
regions. Interestingly, large phase offsets are seen at these
locations, especially at the apex and around the feed-support
legs. The radiation reaching the receiver from these locations is
apparently {\it anti-correlated}, actively diminishing detected signal
strength rather than merely adding incoherent noise.

\begin{figure*}[!t]
  \includegraphics[width=1.0\textwidth]{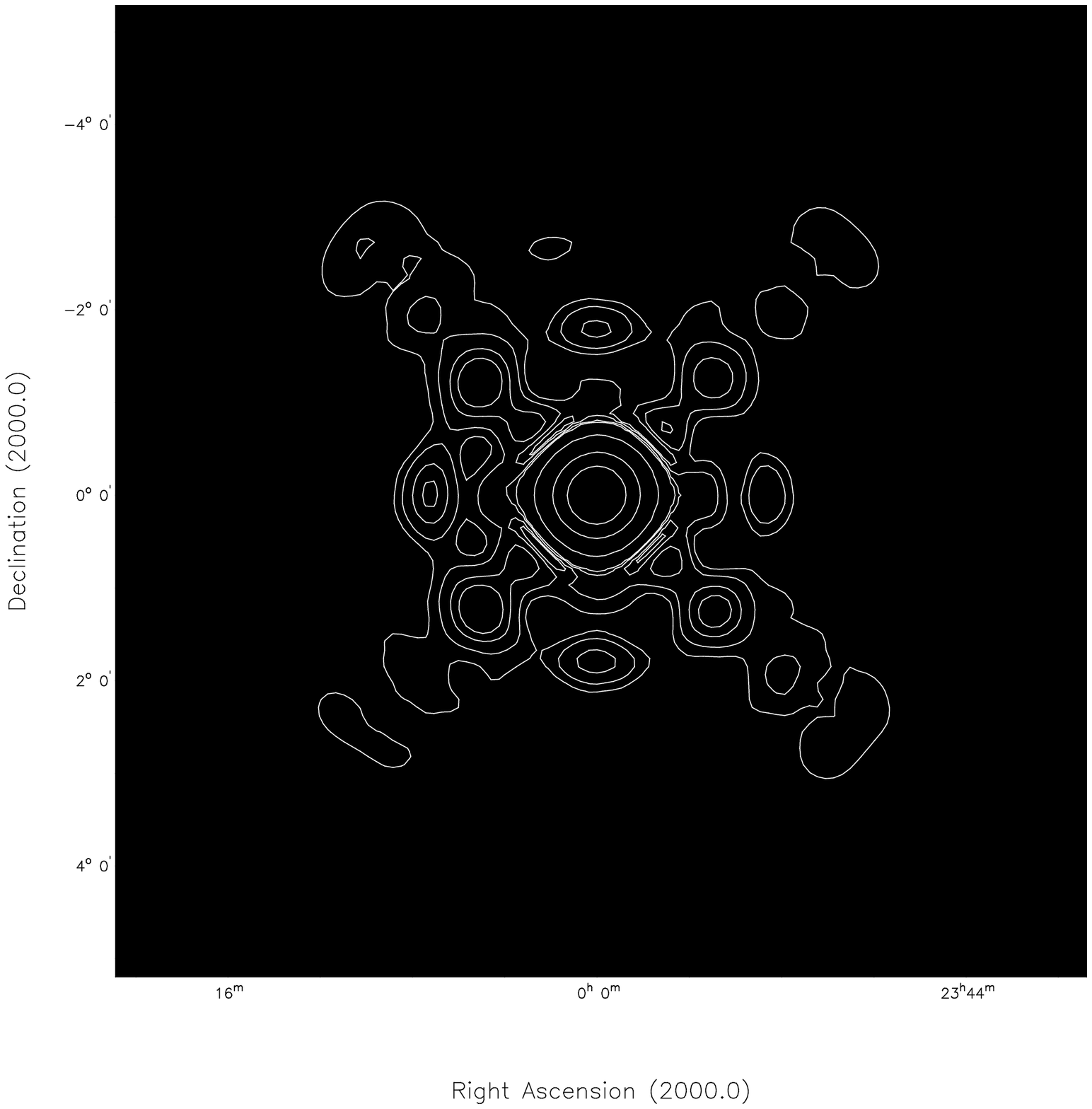}
  \caption{PSF of the beam model averaged over 16.25 MHz (50 channels)
  around 1391.7 MHz. The average of the $XX$ and $YY$ polarisations is
  plotted, corresponding to Stokes {\it I}, such as commonly used for
  correcting the primary beam in reduction software. Contour levels
  are drawn at 0.0005, 0.001, 0.002 0.003, 0.03 0.2 and 0.5, the peak
  value is normalised to 1.}
  \label{beampsf}
\end{figure*}

\subsection{Beam Model Cubes}

Beam models were produced for every pair of observed frequency
channels (yielding 0.625 MHz resolution). The first and last few
channels of each observed 20 MHz band were not used, because of the
roll-off of the bandpass. Given the 8 band centers noted previously,
almost complete frequency coverage between 1322 and 1457 MHz was
available. An interpolation was done between the separate models to
determine a model at each integer MHz within this range. Since voltage
beams were available in the X and the Y polarization, power beams
could be calculated for all four polarization products: \textit{XX,
YY, XY} and \textit{YX}.

\section{Results}
\label{results}

\begin{figure*}[!]
  \includegraphics[width=0.95\textwidth]{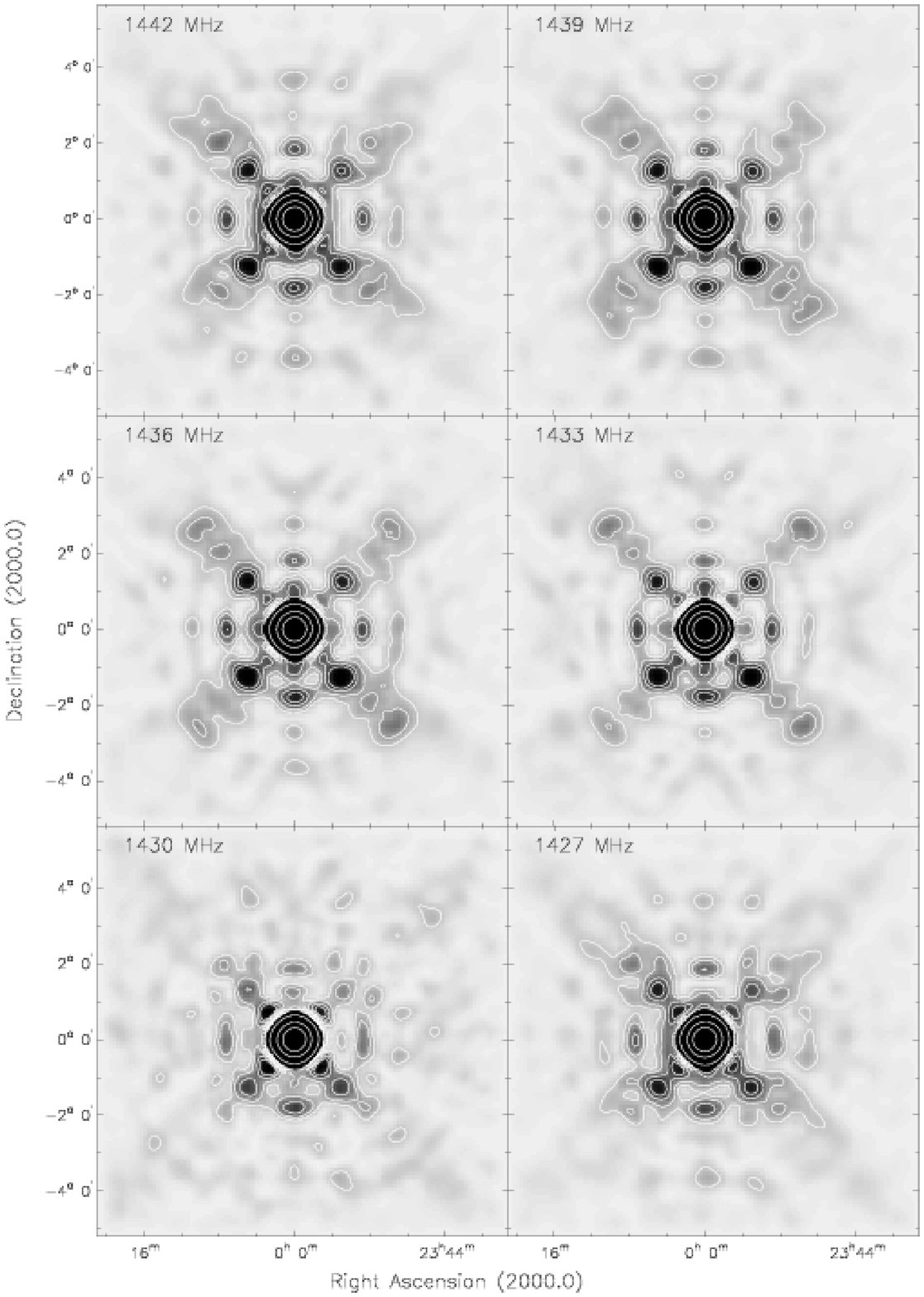}
  \caption{Channel maps of the model cube for Stokes $I$ between 1442
  and 1427 MHz. Clear variations can be seen in the first and second
  side lobe structures. Contour levels are drawn at 0.0005, 0.001,
  0.002, 0.003, 0.03, 0.2 and 0.5.}
  \label{modelmap}
\end{figure*}

By using holographic measurements at different sampling scales, we can
construct an empirical model, that describes the primary beam of the
WSRT with high precision at different frequencies. A Point Spread
Function of the Stokes $I$ polarisation is shown in Fig.~\ref{beampsf}
where the beam is averaged over a bandwidth of about 16 MHz. The main
beam has a clear ``diamond'' shape and around the main beam there is a
sharp dip, corresponding to the first null. Note that this first null
can only be discerned when holographic measurements are sampled at
very high resolution in the image plane (and not the aperture plane as
is usual) together with suitably high frequency resolution. The outer
sidelobes show a clear ``cross'' pattern, due to blockage of the
support structure.

\begin{figure}[]
  \includegraphics[width=0.5\textwidth]{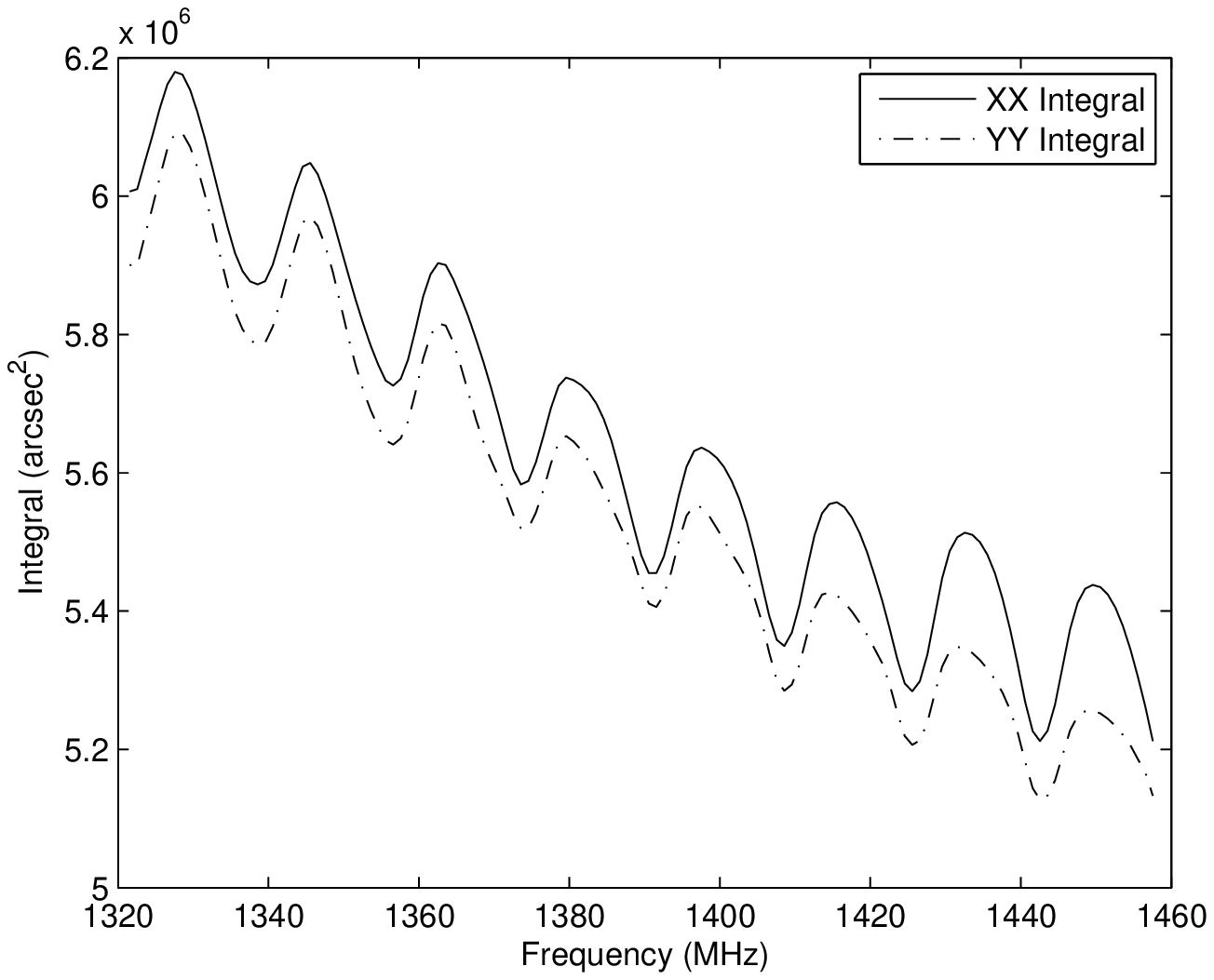}
  \caption{Main beam integral, evaluated from a 2$D$ Gaussian fit for
  the \textit{XX} and \textit{YY} polarizations as function of
  frequency. For increasing frequency the beam integral is decreasing
  quadratically as expected. However, there is a systematic variation
  with a $\sim$17 MHz period.}
  \label{beamintegral}
\end{figure}

Many significant variations can be seen in the model beams as function
of frequency. Channel maps of the Stokes \textit{I}-model are shown in
Fig.~\ref{modelmap}, starting at 1442 MHz, and going down to 1427 MHz
with intervals of 3 MHz. Because of the small scale in this plot,
possible variations in beam-width are difficult to discern
visually. However, there is an easily discernible variation in the
strength of side-lobe structures. For example, at 1433 MHz the first
side lobes are very weak compared to 1430 MHz. On the other hand, the
more distant side lobes almost vanish at 1430 MHz. No extended
sidelobes are seen at 1430 MHz, while at 1439 and 1436 MHz a very
clear diffuse ``cross'' is apparent. The panels in the figure span 15
MHz of bandwidth which is almost one full period in the periodic
variations. These variations are representative of those seen
throughout the full frequency coverage of our model.
There are several ways to visualise the frequency dependence of the
beam and of the side lobes with respect to the main beam. Variations
in the main beam properties can be illustrated by evaluating the beam
integral at all frequencies. This was accomplished with an elliptical
Gaussian fit to each beam using the IMFIT task of the MIRIAD package
\citep{1995ASPC...77..433S}. The values are plotted against frequency
in Fig.~\ref{beamintegral} for both the \textit{XX} and \textit{YY}
polarisations after smoothing over 3 MHz. Very similar patterns are
seen in both cases, although the beam areas for the $XX$ polarisation
are systematically larger by about 2\%.  Similar results are obtained
by simply taking the sum of the relevant image pixels within the main
lobe, although the discrete sampling introduces additional
quantization noise.

As expected, the beam area decreases approximately quadratically with
increasing frequency, since the beamwidth should vary roughly as
$\lambda$/D. More remarkable are the semi-periodic oscillations.  Both
polarizations display a similar periodicity of $\sim$17 MHz with an
amplitude of about $\sim$4 $\%$ in surface area.

\begin{figure}[]
  \includegraphics[width=0.5\textwidth]{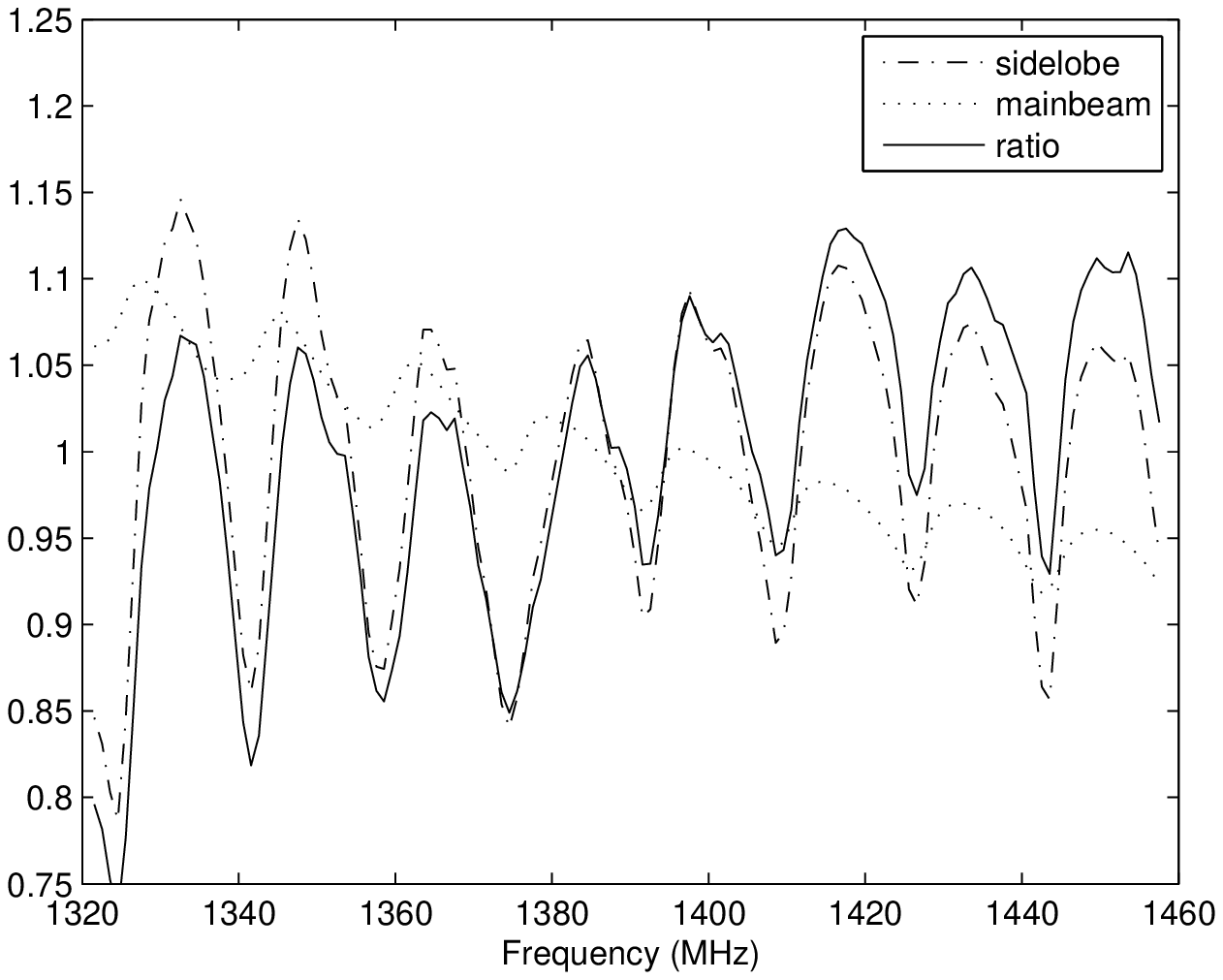}
  \caption{Integrated side-lobe power (dash dotted line), integrated
  main beam power (dotted line) and the side-lobe to main beam power
  ratio (solid line) for Stokes $I$. All three lines are scaled to a
  mean value of one.}
  \label{fitbeam1}
\end{figure}

An estimate of the side-lobe power can be obtained from the difference
of the integral over $11 \times 11^\circ$ with the main-lobe integral
based on the elliptical Gaussian fit as described above. The
integrated main-lobe and side-lobe powers as a function of frequency
are shown for Stokes $I$ after normalisation to a mean value of unity
in Fig.~\ref{fitbeam1} (dotted and dash-dotted line) together with the
ratio between side-lobe and main-lobe power (solid line). This ratio
gives the fraction of the power that is contained in the sidelobes,
independent of any scaling. The same periodicities can be seen as in
the previous plot, for both the main beam and the side lobes, although
the variations in the integrated side lobe power are stronger and are
of the order of $\sim$20~$\%$.

An important aspect of the comparison of beam attributes with
frequency is the normalisation that has been applied. Since the peak
image-plane response of each model beam has been set to unity, no
information is preserved regarding the absolute telescope gain as a
function of frequency. This is important, since normally one might
assume that a smaller main beam area might correspond to a higher
telescope gain, for example via a more uniform aperture illumination
pattern. Such straightforward associations may not
necessarily apply.

\begin{figure}[]
  \includegraphics[width=0.5\textwidth]{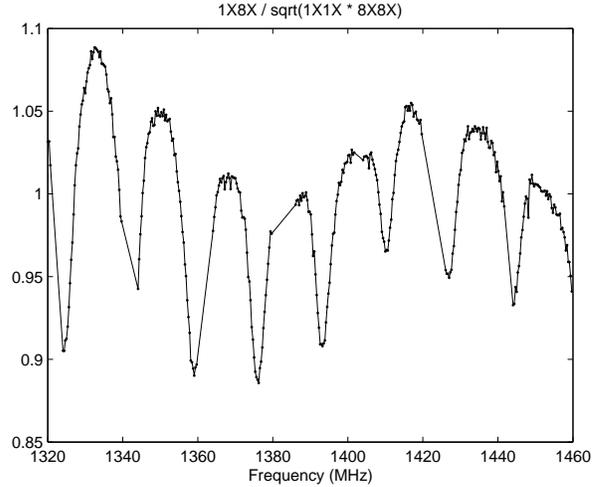}
  \caption{cross-correlations of antenna 1 and 8 divided by the
  auto-correlation of these two antennas. The variation has a period
  of $\sim$17 MHz.}
  \label{ripple}
\end{figure}

\begin{figure}[]
  \includegraphics[width=0.5\textwidth]{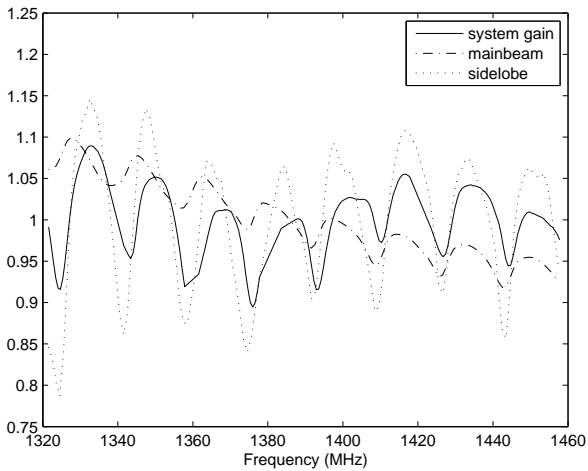}
  \caption{The smoothed gain ripple (solid line), and integrated
  values for the main beam (dash dotted line)}
  \label{gainbeam}
\end{figure}

A convenient method of determining the telescope gain as function of
frequency is available for an interferometer if both auto- and cross-
correlation spectra are measured. This is done by taking the
cross-correlation spectrum of a pair of telescopes, $\rho_{12}(\nu)$,
when observing a bright calibration source and dividing it by the
square root of the product of the two auto-correlation spectra of
these same telescopes, $\rho_{11}(\nu)$ and $\rho_{22}(\nu)$. In terms
of the effective aperture $A$, the source flux density, $S$, and the
bandpass shape, $F$, one gets,
\begin{equation}
 \rho_{12}(\nu) \propto {A(\nu) \cdot S(\nu) \cdot F_1(\nu) \cdot
  F_2(\nu) \over [T_{sys} + T_{sou}(\nu)]}
\end{equation}
and
\begin{equation} 
\rho_{11}(\nu) \propto {[T_{sys} + T_{sou}(\nu)]}\cdot F_1(\nu) \cdot
F_1(\nu) 
\end{equation}
so
\begin{equation}
{\rho_{12}(\nu) \over \sqrt{ \rho_{11}(\nu) \cdot \rho_{22}(\nu)}}
  \propto {A(\nu) \cdot S(\nu) \over [T_{sys} + T_{sou}(\nu)]^2}
\end{equation}
where the system temperature is made up of a source independent
component, $T_{sys}$ and the contribution due to the observed source,
$T_{sou}(\nu)$.  The bandpass shape of both receiver chains cancels
out in this combination, and one is left with the gain spectrum of the
interferometer (together with possible frequency structure of the
calibration source). The receiver temperature due to the observed
source will in turn be proportional to $A(\nu)$, so if this were
neglected it would lead to a dilution of any possibile variation of
effective aperture with frequency. In practise, 3C147 increases the
system temperature by only 8\% for the WSRT system near 1400~MHz, so
this dilution effect could be compensated by simply multiplying the
deduced gain variations by a factor of 1.002. The derived gain is
plotted for two representative telescopes (RT1 and RT8) in the
\textit{X} polarization in Fig.~\ref{ripple} using one of our
holographic observations of 3C147 during an interval when both the
reference and scanning telescopes are pointed at the source. Very
substantial oscillations of 5 to 10\% amplitude are seen with a
similar period of $\sim$17 MHz. Comparing the system gain with the
main and side-lobe powers in Fig.~\ref{gainbeam} illustrates the
counter-intuitive result that the highest gain is accompanied by both
the largest beam area and the highest ratios of side- to main-lobe
power.

One way in which to partially understand these correspondences is to
consider that the broadest main-beam patterns may correspond to the
instances of highest effective illumination of the central portions of
the aperture with respect to the outer portions of the aperture. This
might occur when a minimum amount of radiation (with inappropriate
phase) reaches the feed after reflection from the central blockage
(in the shadow of the receiver package). At other portions of the
cycle, a larger amount of radiation from this region would reach the
feed, causing both a decrease in gain (in view of the detrimental
phase) as well as a narrower main beam, since the central portion of
the reflector would contribute negatively. What remains difficult to
explain, is why the side-lobe levels would be lowest when the gain was
lowest.

\section{Testing with Celestial Sources}
\label{testing}

\begin{table*}[!t]
\begin{tabular}{lllllll}
\hline
\hline
Source  & Right Ascension (hms) & Declination (dms) & Radius (arcsec) & cos$^6$ flux (Jy) & model flux (Jy) & NVSS flux (Jy)\\
\hline
A  &    8:21:29   &    +70:14:33   &     1904   &    115.3   &    107.7  &   101.8  \\ 
B  &    8:22:16   &    +70:53:08   &     1124   &    613.5   &    605.4  &   602.3  \\ 
C  &    8:21:33   &    +71:07:43   &     1658   &    125.3   &    125.4  &   116.3  \\ 
D  &    8:21:31   &    +71:19:41   &     2309   &    975.0   &    1012.5  &   961.9  \\ 
\hline
\hline
\end{tabular}
\caption{Data for some extreme off-axis sources found in the Holmberg
II field. The distance from the primary beam center is given in the
fourth column. The fifth and sixth column give the average flux of all
the narrow band images (for an effective frequency of 1360 MHz) after
primary beam correction with both the cos$^6$ beam description and the new
empirical model. The NVSS flux at 1400 MHz is given in the last
column.}
\label{hoiitable}
\end{table*}

\begin{figure*}[!t]
  \includegraphics[width=1.0\textwidth]{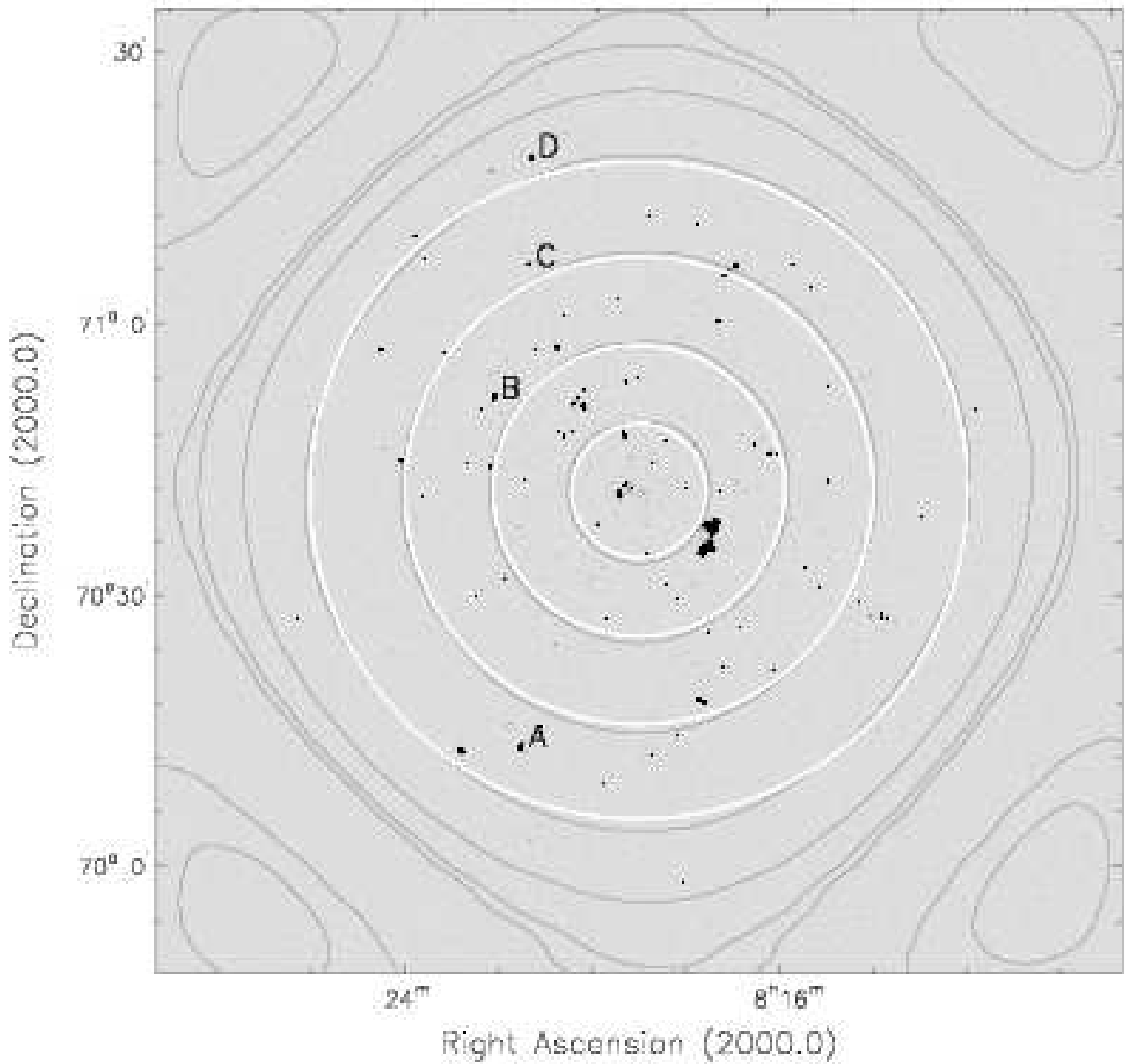}
  \caption{WSRT image of the Ho~II field. The white contours represent
  the old cos$^6$ beam description, the contour levels from inside out
  are at 0.9, 0.5, 0.2 and 0.05. The gray contours correspond to the
  new model, the levels are at 0.9, 0.5, 0.2, 0.05, 0.01, 0.003 and
  0.002. In the corners of the field, the first sidelobes can be
  clearly recognized. Letters are assigned to those sources for which
  we show the beam-corrected flux as function of frequency in
  Fig.~\ref{nvssABCD}.}
  \label{hoiibeam}
\end{figure*}

\begin{figure*}[!t]
  \includegraphics[width=0.5\textwidth]{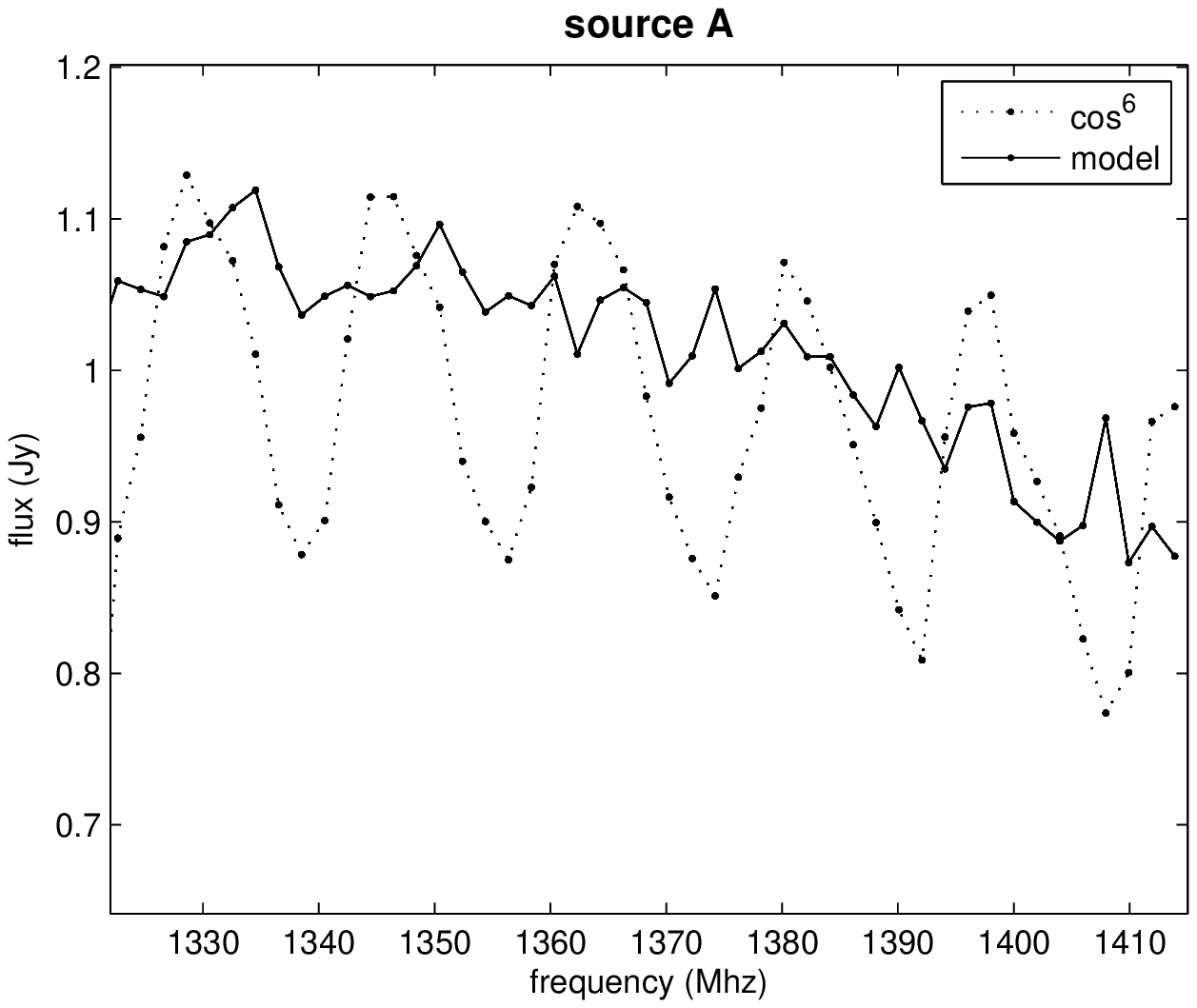}
  \includegraphics[width=0.5\textwidth]{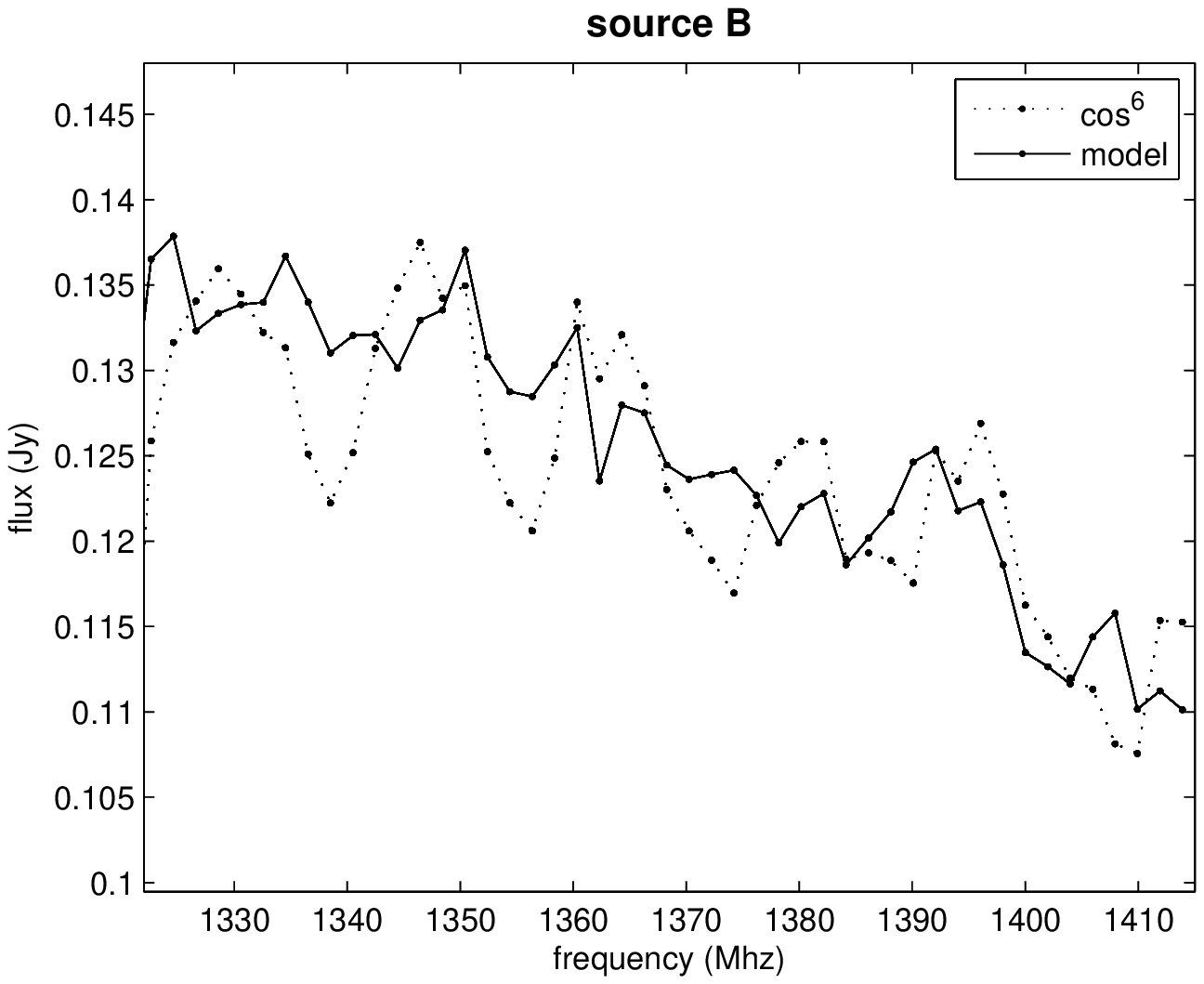}
  \includegraphics[width=0.5\textwidth]{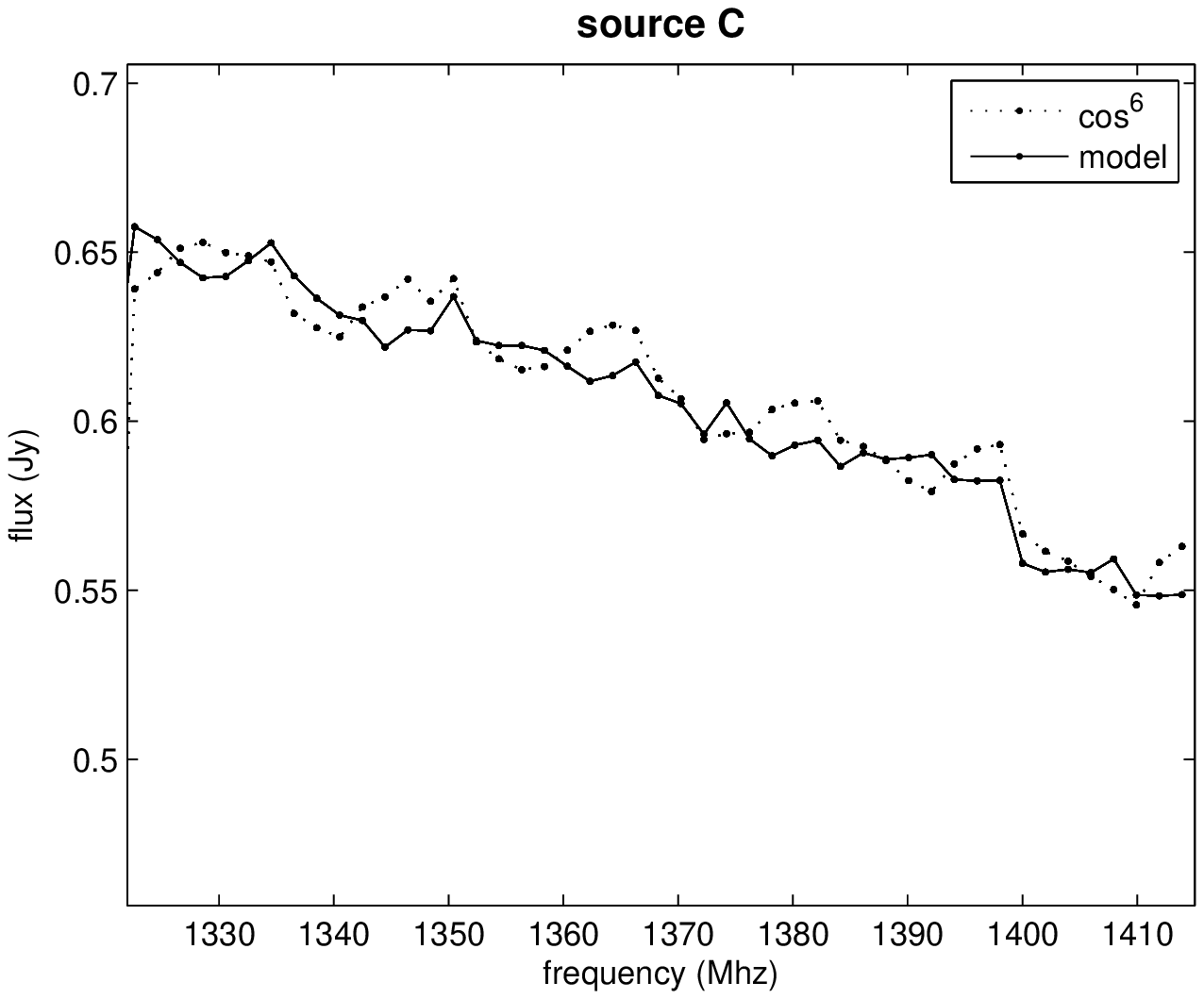}
  \includegraphics[width=0.5\textwidth]{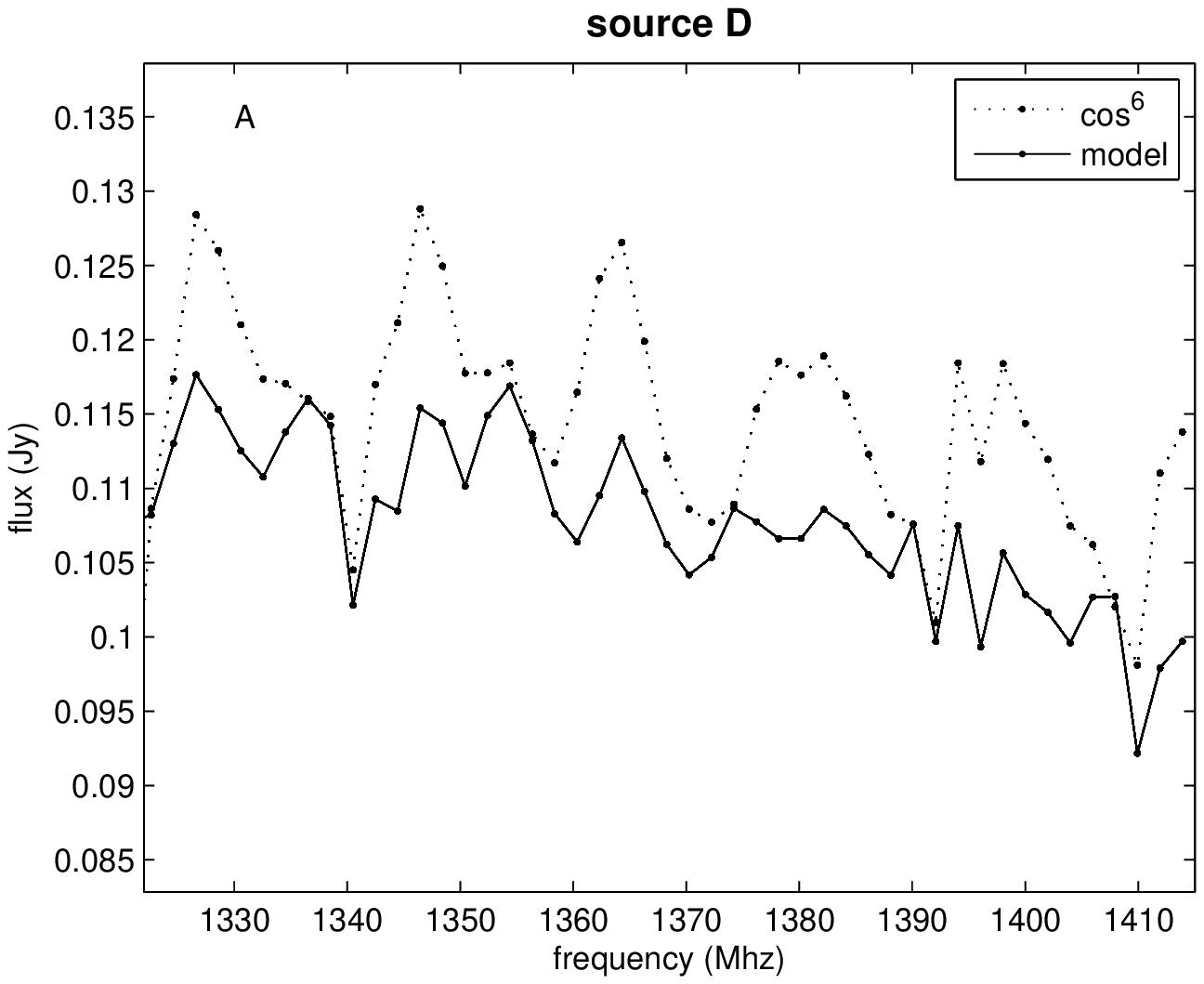}
  \caption{Fluxes are plotted as function of frequency for sources in
  Table ~\ref{hoiitable} after primary beam correction. The solid lines
  correspond to a correction with our empirical model, while the
  dotted lines are corrected using the cos$^6$ beam approximation.}
  \label{nvssABCD}
\end{figure*}

Although we have confidence in our methodology, it is important to
verify the rather remarkable variations in beam parameters with
frequency which we derive, using independent observations under typical
observing conditions. 

A suitable source of single field observations is the recently
published WSRT SINGS survey \cite{2007A&A...461..455B}. We have
produced calibrated, deconvolved images from this survey database for
four different fields centered on the nearby galaxies Holmberg~II,
IC~2574, NGC~2841 and NGC~5194 at frequencies between 1322 and 1408
MHz at increments of 2 MHz. Unresolved and unconfused sources were
sought in each of these fields with a high enough apparent brightness
that they could be detected with good signal-to-noise in the narrow
bandwidth images. The most interesting sources are those that are
bright, but very far from the pointing center since they probe the
largest primary beam corrections. We overlay the cos$^6$ and our
empirical primary beam attenuation model on the field of Ho~II in
Fig.~\ref{hoiibeam} to illustrate how the apparent density of sources
varies with position. Comparison of contours from the two types of
beam model suggests that small departures from axi-symmetry already
set in at relatively high levels. The best test sources found in the
four different fields are listed in Table~\ref{hoiitable}. We also
tabulate the NVSS (NRAO VLA Sky Survey, \cite{1998AJ....115.1693C})
1.4 GHz flux densities of these sources together with the averaged
(ie. 1360 MHz) flux density obtained from both the cos$^6$ function
and our empirical model. Reasonable agreement can be found between the
flux values, although the fluxes we measure at 1360 MHz are
systematically higher than seen in the NVSS (at effective frequency of
1400 MHz) likely due to the negative mean spectral index of background
sources.

\begin{figure*}[!]
  \includegraphics[width=0.5\textwidth]{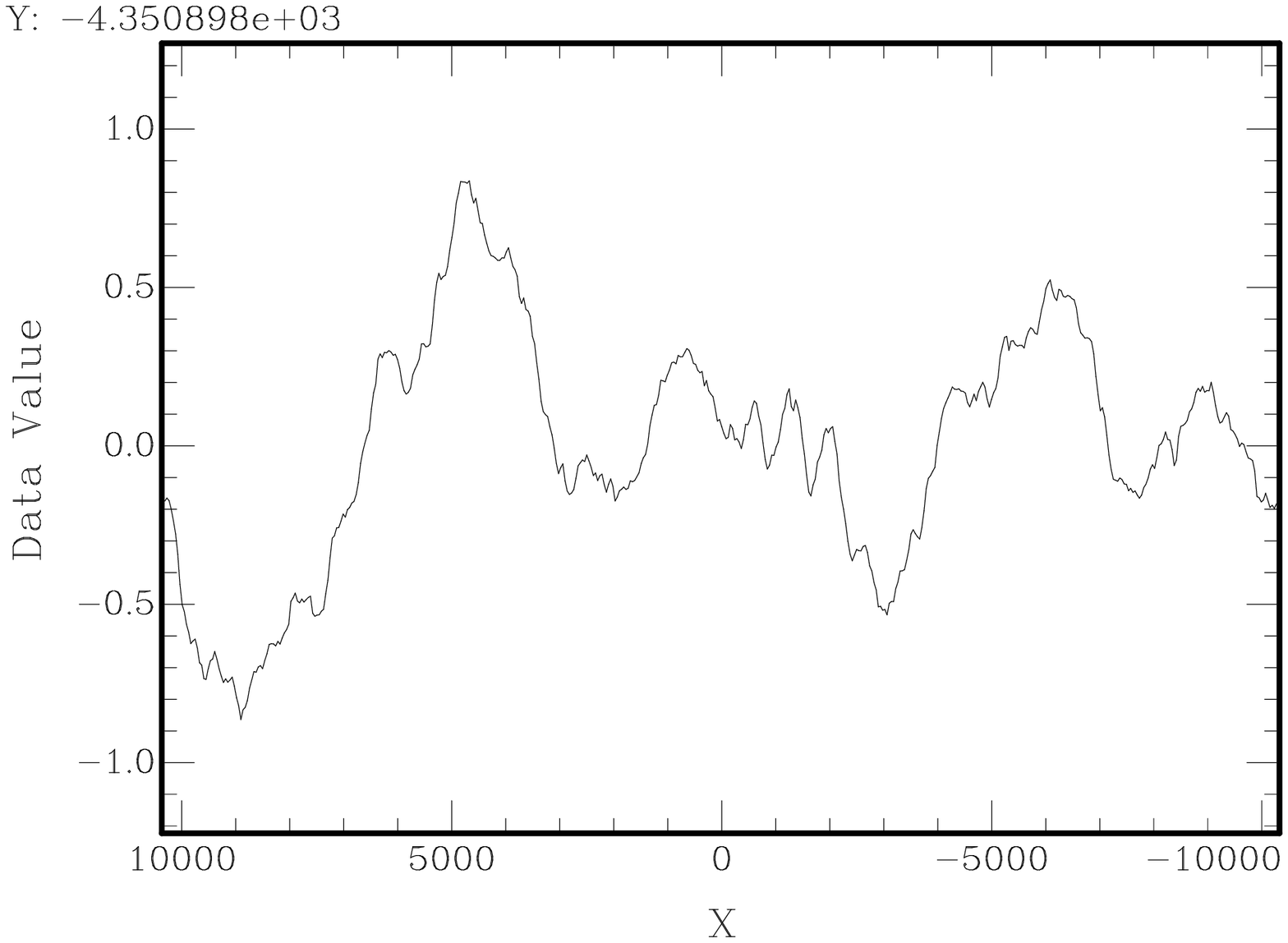}
  \includegraphics[width=0.5\textwidth]{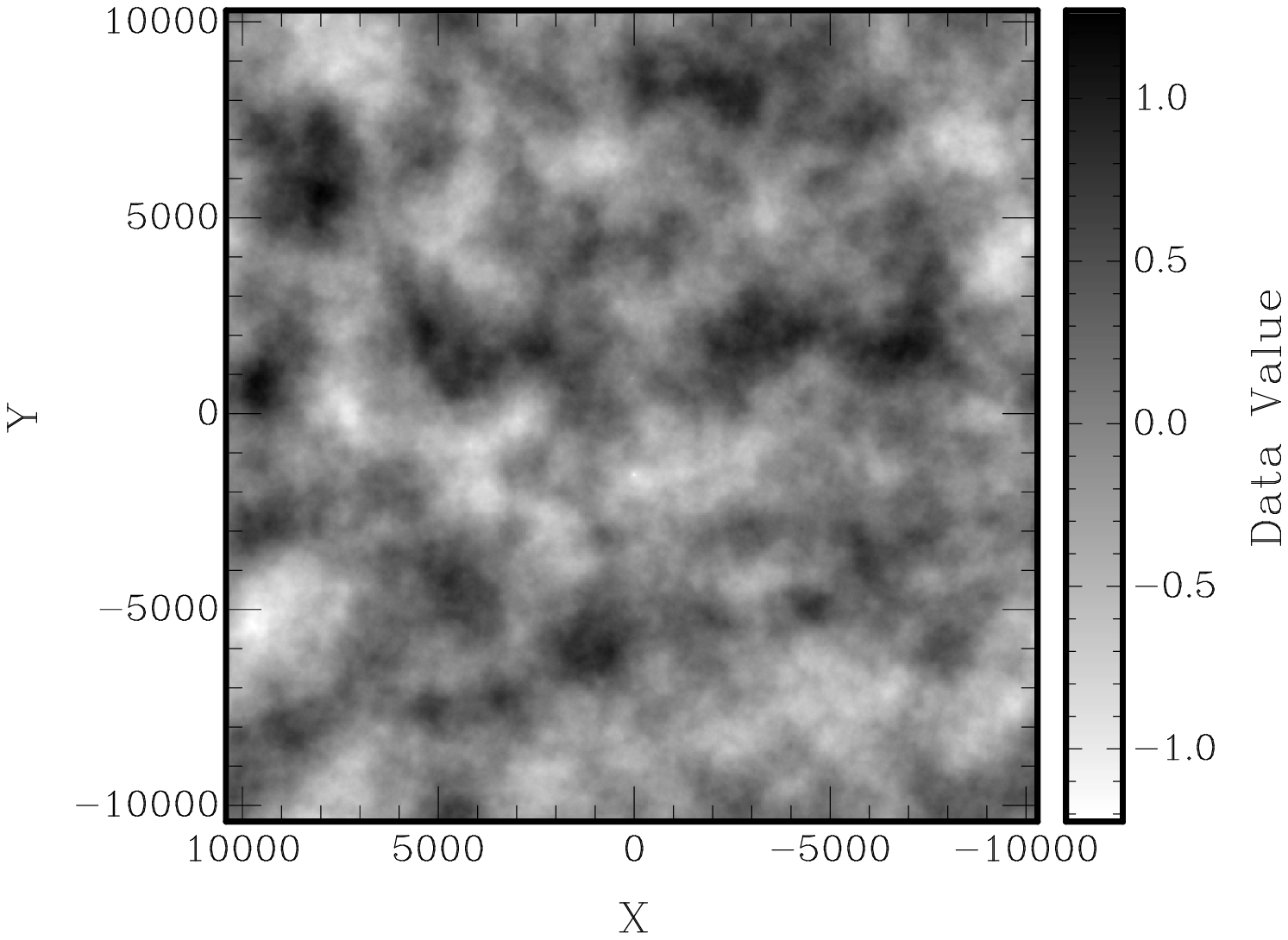}
  \caption{Trial design for an isotropic reflective surface
  treatment. The image at right represents a white noise field after
  tapering with a $-$2 power-law of radius (truncated at 5 pixels from
  the origin) and Fourier transformation. At left is a representative
  cross-cut of the surface to better illustrate the range of
  structures present. To provide performance between observing
  wavelengths of 70 and 1.5~cm the surface should have spatial
  dimensions of about 3.5~m on a side and maximum peak-to-peak
  excursions of about 70~cm.}
  \label{isotile}
\end{figure*}

We plot the primary beam corrected flux densities of several test
sources in Fig.~\ref{nvssABCD}. In the figures we
contrast the flux densities obtained by applying the cos$^6(c \nu r)$
model and our empirical primary beam attenuation model. The
improvement in recovering a plausible, smoothly varying spectrum (and
of course the correct mean flux density) for the off-axis sources is
striking. In the case of source A, there
is a very strong variation in the flux with frequency. When using the
new model, the 17 MHz variation is flattened out and only smaller
noisy fluctuations remain. A quite similar effect is apparent in
source B and source D. Of course the new model does
not predict substantial differences from the historical
analytic model at all positions within the beam. An example is given
with source C, where mean flux levels are comparable, although
the low level periodic variation is eliminated with the new model.


\section{Summary and Discussion}
\label{summary}

In this paper we develop a frequency-resolved empirical model of the
primary beam of the WSRT 25m antennas between 1320 and 1460 MHz in two
orthogonal linear polarizations. This is based on a series of
holographic measurements sampling a wide range of angular scales from
about five arcmin to 11 degrees. Not surprisingly, there are
significant departures from axi-symmetry in the form of a ``diamond''
distortion of the main lobe due to the quadrapod blockage of the
telescope aperture. This aperture blockage also gives rise to a
four-fold symmetry in the side-lobe pattern within a radius of about
five degrees. Systematic differences are also apparent in the two
perpendicular polarizations, in the sense that beam areas are about
2\% larger in the {\it XX} polarization than the {\it YY}. Furthermore
the beam shapes are not symmetric, but there is a slight ellipticity;
elongated in the direction of the polarization. For the {\it XX}
polarization, the beam is about 4\% larger in the x-direction,
compared to the perpendicular y-direction. For the {\it YY} polarization we
see an elongation in the y-direction, although here the difference
is only about 2.5\% relative to x.

More surprisingly, we document systematic oscillations in the beam
properties as a function of frequency. A similar semi-sinusoidal
variation is found for each of: (1) the integrated beam area, (2) the
ratio of integrated side-lobe to main-lobe power and (3) the effective
aperture of the telescope system. All three of these attributes are
modulated at the 5 to 10\% level with a basic periodicity of about
17~MHz, corresponding to the natural ``standing wave'' period of about
$c/2f$ for a paraboloid of focal distance, $f=8.75$ m. These effects are
manifestations of constructive and destructive interference ocurring
at the telescope feed caused by multi-path transmission. 

We have verified these deduced oscillations in beam properties with
independent measurements under typical observing conditions; namely
long duration tracks employing earth rotation synthesis. A crucial
aspect here is that the WSRT telescopes employ an equatorial mount, so
that the beam patterns are at least fixed on the sky despite tracking
a field for many hours. Although the on-axis frequency response has
been effectively calibrated in such observations, continuum sources
which are sufficiently far off-axis display a modulation in their
apparent flux density with frequency.  Application of our empirical
beam model effectively removes such off-axis modulation of apparent
flux density with frequency. While this procedure permits effective
correction of interferometric observations (employing an equatorial
mount), albeit at a substantially increased complexity of processing,
there is no comparable correction procedure for total power
measurements which suffer from the same effect. Improving the spectral
baselines of total power observations will require a method of
physically eliminating such modulations from the telescope/feed system.

The primary surfaces which contribute to these interference effects
are those portions of the main reflector which suffer from geometrical
blockage; the regions ``under'' the feed-support legs and prime focus
receiver support structure (as seen from the sky) together with those
parts of the reflector which are ``behind'' the feed-support legs as
seen from the feed. And finally, the surfaces of the feed-support legs
themselves which are directly visible from the feed. Since coherent
reflections from these surfaces appear to be responsible for the
oscillations in antenna beam properties with frequency we suggest that
a likely remedy for these effects would be a surface treatment of all
of the aforementioned surfaces to make them behave as broad-band
isotropic scatterers rather than coherent reflectors. Scattering is
prefered over absorption since large absorbing surfaces at ambient
temperature on the reflector and telescope structure will adversely
affect the system temperature.

Design of a broad-band isotropic reflector could be achieved by taking
a white noise distribution, applying a suitable power-law of radius taper and
Fourier transforming this distribution. A logical choice for the
power-law index might be $-$2 which would yield an equal amount of
integrated power in surface irregularities on all spatial scales (in two
dimensions). The range of spatial scales over which the power-law
extends would determine the band-width over which the surface should
function. The peak-to-peak surface excursions should correspond to the
maximum wavelength for which the surface should be effective. 

We have produced a trial design for an isotropic reflective surface
intended to be effective between wavelengths of about 70~cm and
1.5~cm, which we illustrate in Fig~\ref{isotile}. The $-$2 power-law
taper that was applied to the white noise field has been truncated at
a radius of 5 pixels so as to provide about five by five of the
largest scale fluctuations in the surface after Fourier
transformation. The physical dimensions of the depicted surface would
be about 3.5~m on a side with maximum peak-to-peak fluctuations of
about 70~cm. The power-law extends down to size scales of 0.7~cm which
should still provide the desired scattering properties at 1.5~cm
wavelength. If regions larger than 3.5~m were being equipped with such
a surface treatment they would ideally be taken from a larger input
noise field, to minimize edge effects between ``tiles'' as well as any
repetition of the tiling pattern (that might introduce grating-like
responses in special directions).

We have embarked on fabrication of a scale-model of the illustrated
surface design, which will be subject to measurement in a test
range. If the expected broad-band scattering properties are confirmed,
we will pursue full scale deployment of this surface treatment on the
Parkes telescope. Perhaps the era of ``standing waves'' limiting
performance in radio telescopes is approaching an end.

\begin{acknowledgements}
We would like to thank Hans van Someren Greve for developing software
for holographic data reduction (MSHOLOG) and adjusting it for our
purposes. We made extensive use of this tool. Furthermore we thank
John Romein for adding support for empirical beam models into several
tasks in the ``Miriad'' package. This permitted us to efficiently test
the application of our model to real images. We are grateful to
Michael Kesteven for useful comments on the original manuscript. The
Westerbork Synthesis Radio Telescope is operated by the ASTRON
(Netherlands Foundation for Research in Astronomy) with support from
the Netherlands Foundation for Scientific Research (NWO).
\end{acknowledgements}

\bibliographystyle{aa}
\bibliography{names,bibliography}
\end{document}